\documentclass[11pt,reqno]{amsproc}
\usepackage{fullpage}
\usepackage{tikz,everypage}
\usepackage[semicolon,square,authoryear]{natbib}
\numberwithin{equation}{section}
\usepackage{cite}
\usepackage{color}
\usepackage{graphicx}
\usepackage{subfig}
\usepackage{wrap fig,lip sum,booktabs}
\usepackage{caption}
\usepackage{comment}
\usepackage{multirow}
\usepackage{rotating}
\usepackage{epstopdf}
\usepackage{enumerate}
\usepackage{enumitem}
\usepackage[debug=false, colorlinks=true, pdfstartview=FitV, linkcolor=blue, citecolor=blue, urlcolor=blue]{hyperref}
\newtheorem{theorem}{Theorem}

\newtheorem{remark}{Remark}
\newtheorem{definition}[theorem]{Definition}

\newlength{\drop}
\definecolor{amethyst}{rgb}{0.6, 0.4, 0.8}
\definecolor{burgundy}{rgb}{0.5, 0.0, 0.13}

\title{A performance spectrum for parallel computational frameworks that solve PDEs}

\author{\textbf{J.~Chang}, 
\textbf{K.~B.~Nakshatrala}, 
\textbf{M.~G.~Knepley} and 
\textbf{L.~Johnsson}\\
  {\small Correspondence to: \textbf{\emph{e-mail:}} knakshatrala@uh.edu,
  \textbf{\emph{phone:}}+1-713-743-4418}} 

\date{\today}
\begin{document}


\begin{titlepage}
    \drop=0.05\textheight
    \centering
    \vspace*{0.3\baselineskip}
    \rule{\textwidth}{1.6pt}\vspace*{-\baselineskip}\vspace*{2pt}
    \rule{\textwidth}{0.4pt}\\[0.3\baselineskip]
    {\LARGE \textbf{\color{burgundy} 
   A performance spectrum for parallel 
   computational \\[0.3\baselineskip] 
   frameworks that solve PDEs}}
   \\[0.3\baselineskip]
   \rule{\textwidth}{0.4pt}\vspace*{-\baselineskip}\vspace{3.2pt}
    \rule{\textwidth}{1.6pt}\\[0.3\baselineskip]
    \scshape
     An e-print of the paper is available 
     on arXiv. \par
    \vspace*{0.3\baselineskip}
%
    {\Large J.~Chang\par}
    {\itshape Graduate Student, University of Houston.}\\[0.3\baselineskip]
{\Large K.~B.~Nakshatrala\par}
    {\itshape Department of Civil \& Environmental Engineering \\
      University of Houston, Texas 77204--4003.\\
    \textbf{phone:} +1-713-743-4418, \textbf{e-mail:} knakshatrala@uh.edu \\
    \textbf{website:} http://www.cive.uh.edu/faculty/nakshatrala}\\[0.3\baselineskip]
    {\Large M.~G.~Knepley\par}
    {\itshape Department of Computational 
    and Applied Mathematics, 
      Rice University.}\\[0.3\baselineskip]
    {\Large L.~Johnsson\par}
    {\itshape Department of Computer Science, 
      University of Houston.}
\vspace{-0.2in}
\begin{figure*}[h]
\subfloat{\includegraphics[scale=0.82]{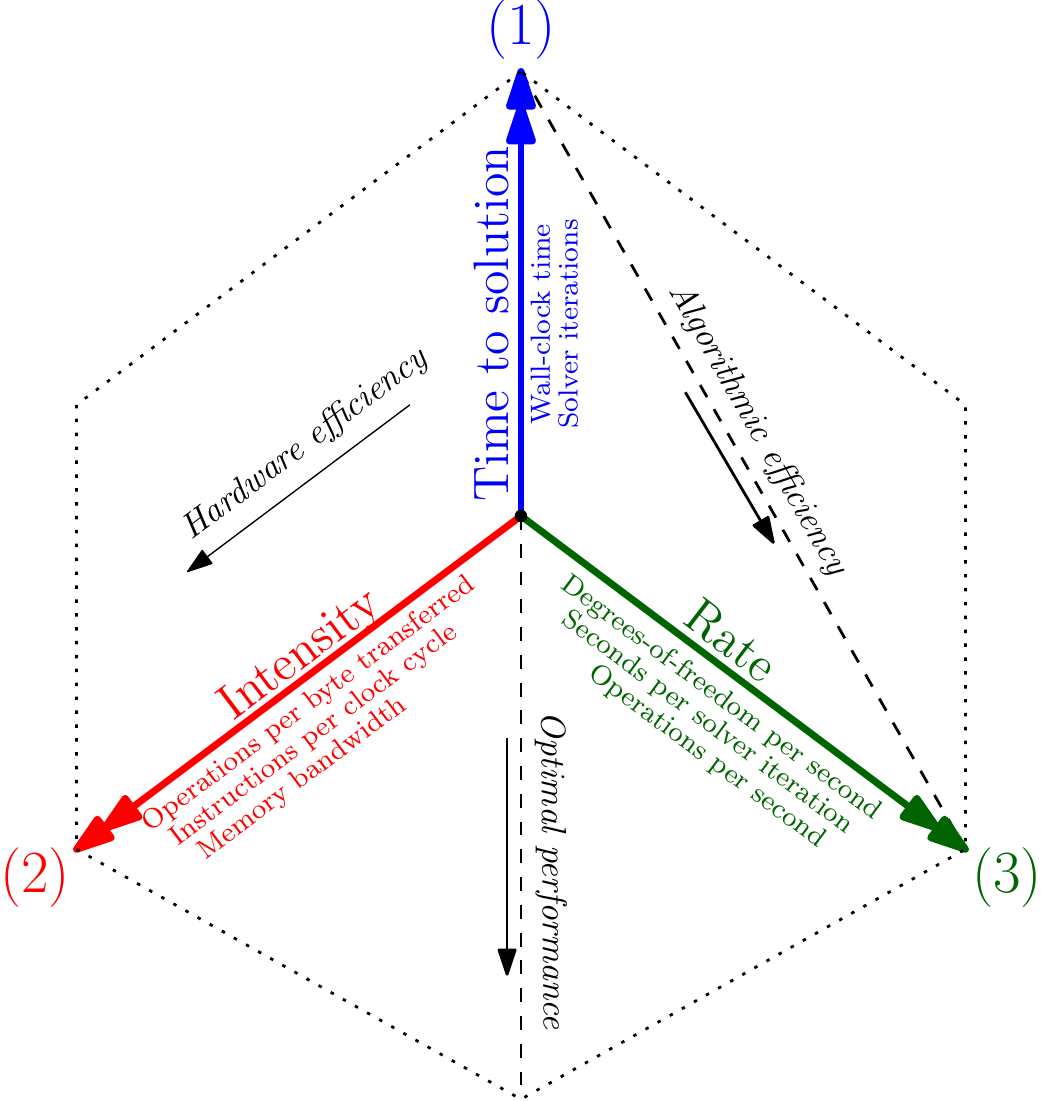}}
\captionsetup{format=hang}
\vspace{-0.12in}
\caption*{\emph{Proposed performance 
spectrum that documents time, intensity 
and rate.}}
\end{figure*}
\vfill
{\scshape 2017} \\
{\small Computational \& Applied Mechanics Laboratory} \par
\end{titlepage}

\begin{abstract}
  Important computational physics problems are often
  large-scale in nature, and it is highly desirable
  to have robust and high performing computational
  frameworks that can quickly address these problems.
  However, it is no trivial task to determine whether
  a computational framework is performing efficiently
  or is scalable. 
The aim of this paper is to present various strategies for better understanding the 
performance of any parallel computational frameworks for solving PDEs. 
Important performance issues that negatively impact time-to-solution are 
discussed, and we propose a performance spectrum analysis that can enhance one's 
understanding of critical aforementioned performance issues. As proof of 
concept, we examine commonly used finite element simulation packages and software 
and apply the performance spectrum to quickly analyze the performance and 
scalability across various hardware platforms, software implementations, 
and numerical discretizations. It is shown that the proposed performance spectrum is 
a versatile performance model that is not only extendable to more complex PDEs 
such as hydrostatic ice sheet flow equations, but also useful for understanding
hardware performance in a massively parallel computing environment. 
Potential applications and future extensions of this work are also discussed.
\end{abstract}

\maketitle

\section{INTRODUCTION AND MOTIVATION}
\label{Sec:Introduction}
Both efficient algorithms and software performing well on modern computing
systems are crucial to address
current scientific and engineering problems. These
tools are important for bridging the gap between theory and 
real-world data. Such problems often need to tackle field-scale data
sets using parallel computers, parallel algorithms, and programming
tools such as OpenMP \citep{dagum1998openmp} and the Message Passing 
Interface (MPI) \citep{gropp1999using} and cannot be solved on a 
standard laptop or desktop. For example, hydrologists and geophysicists 
need to work with field-scale reservoirs which could span tens of 
kilometers and evolve on time scales of hundreds of 
years. Moreover, such reservoir simulations involve complicated multi-phase 
and multi-component flows which require multiple complex equations to 
be solved accurately and efficiently. Atmospheric and climate modelers
also require state-of-the-art techniques as both data assimilation and
parameter estimation need to be performed quickly on meso-scale and 
global-scale applications. The US Department of Energy has 
invested in the development of several portable and extensible 
scientific software packages like PETSc \citep{petsc-user-ref,petsc-web-page} 
and PFLOTRAN \citep{pflotran-web-page} that can help address 
such important large-scale problems. The time spent developing 
parallel computational frameworks is amortized when application 
scientists employ the packages in their work.

However, it is not always known whether the performance of a particular 
parallel computational framework or software will be satisfactory across a panoply of 
solvers and computing platforms. How can one really tell whether an algorithm 
is performing at its highest level? Is there room for improvement? Answering 
these questions in full is a Herculean task, but questions regarding the 
algorithmic and computational efficiency of scientific tools and 
libraries still need to be answered \citep{keyes2013multiphysics}. Hence,
we need \emph{performance models} which enable us to synthesize 
performance data into an understandable framework. Performance models
can include many metrics of importance such as total floating point 
operations (FLOP), memory usage, inter/intra process/node communication,
memory/cache bandwidth, and cache misses/hits. If not carefully optimized, 
some of the hardware resources can become unnecessary bottlenecks that result in costly and 
inefficient numerical simulations. Modern computer systems are quite
complex and the performance can be difficult to predict with good
accuracy. Conducting large-scale simulations on state-of-the-art 
supercomputers may require hundreds to thousands of hours of compute time, 
so it is highly desirable to have a performance model that can predict 
how a particular parallel computational framework may perform. The 
application or domain scientist may use software that either is
not made in house or is a ``black-box'' tool, and it would be too
time consuming, or impossible if source code is unavailable, to 
dissect the code and analyze the 
design of the subroutines and data structures. It is therefore 
desirable to analyze these codes as a whole.
\subsection{Review of previous works}
We now briefly highlight some useful approaches and models one 
could take to analyze and perhaps improve the performance of 
any parallel computational framework. One of the simplest measures 
one can utilize is the STREAM memory-bandwidth benchmark \citep{streams}. 
This benchmark measures sustainable memory-bandwidth on a single server 
and indicates the number of threads that saturates memory bandwidth.
Memory-bandwidth is an important limitation 
to consider on modern machines \citep{wulf1995hitting,mckee2004reflections,
murphy2007effects}

The Roofline model \citep{Williams_ACM_2009,Lo_roofline} captures peak 
achievable performance on a server taking into account both CPU and 
memory-bandwidth capabilities by introducing the Arithmetic Intensity (AI). The 
AI is simply the measure of the total floating-point operations needed, total
FLOP, over Total Bytes Transferred (TBT).
Higher AI's indicate that the algorithm or computational framework is
more computationally intensive and requires less bandwidth for a given
amount of work. One is free to employ any cache model
when determining the TBT metric for the roofline model. 
For example, scientists have developed a Sparse Matrix-Vector (SpMV) 
multiplications model \citep{Kaushik99towardrealistic} which is based on 
``perfect cache" (i.e., matrices and vectors are loaded and 
stored once from memory). SpMV is an integral part 
of iterative solvers for solving PDEs. It has been 
shown in \citep{Chang_JOMP_2016} that the SpMV ``perfect cache" model
can also be used to accurately predict and understand the hardware 
performance of optimization-based solvers for enforcing discrete 
maximum principles. In \citep{may_ptatin}, the authors employ 
matrix-free iterative methods for Stokes equation, which is needed
for lithospheric dynamic applications. The authors manually count the TBT 
based on source code. The advantage of matrix-free methods is that the sparse
matrix-vector multiplication, which is memory-bandwidth limited, is not explicitly
stored thus bringing the computational frameworks' upper-bound limit 
of the roofline closer to the Theoretical Peak Performance (TPP) region.
TBT can also be determined based on memory level traffic or cache misses. 
The same analysis can be carried out for many-core architectures,
such as Nvidia GPUs and the Intel Xeon Phi ``Knights Landing'' (KNL), 
in \citep{KnepleyTerrel2013,KnepleyRuppTerrel2016}.

For a more thorough analysis of performance, advanced software tools such as
the HPCToolkit \citep{adhianto2010hpctoolkit}, OpenSpeedShop \citep{schulz2008open},
{Scalasca \citep{geimer2010scalasca}, and TAU \citep{shende2006tau}}
are used by scientific software developers and application scientists alike.
These tools provide in-depth performance analyses 
of scientific codes and can also be used to debug the codes. Many of them
also rely on PAPI \citep{mucci1999papi} 
which use low level hardware counters for important metrics like 
FLOPs, total CPU (central processing unit) cycles, and 
cache misses. These tools have proven to be extremely useful for 
computational scientists in all areas of computational 
physics and can provide a good understanding of the hardware 
performance of any computational framework for solving PDEs.

\subsection{Main contributions}
In this paper, we provide a simple and easy-to-use performance
model that can be used in addition to the techniques and tools
mentioned above. Our performance model, which we refer to as a
\emph{performance spectrum}\footnote{We borrowed the terminology ``\emph{performance
    spectrum}'' from Dr.~Jed Brown, which he used in 
  his presentation at 2016 SIAM Parallel Processing
  conference, which was held at Paris, France
  (conference website:~http://www.siam.org/meetings/pp16/).}
takes into account time-to-solution, AI based on cache misses, and 
equations solved per second. This model is applicable to any level of 
a scientific code, whether it be the entire computational framework or 
only particular phases or functions such as mesh generation, assembly of a
matrix, or the solver step. \emph{It is important to note that this tool is 
not intended to replace any of the aforementioned performance 
tools or models but to simply augment one's ability to quickly understand 
and diagnose the performance from both the hardware, software, and algorithmic 
stand point.} The main contributions of this paper can be enumerated as follows: 
\begin{enumerate}
\item We outline common issues pertaining to performance,
  ways to identify them, and methods to address them.
\item We present a model called performance spectrum that provides
  an enhanced understanding of the performance and
  scalability of algorithms and software.
\item We demonstrate that the proposed model 
  can be utilized on existing popular software
  packages and solvers.
\item We apply the model to a more complicated and 
  nonlinear PDE and document the parallel performance 
  of the computational framework across HPC machines.
\item We discuss some possible ways in which this
  performance spectrum model can be extended.
\end{enumerate}

The rest of the paper is organized as follows. 
In Section \ref{Sec:Issues}, we outline some of the key performance
issues one may come across when solving PDEs
and how to address some of them. In Section
\ref{Sec:Model}, we propose a model, performance spectrum, which captures 
three critical metrics useful for understanding 
performance and scalability. In Section \ref{Sec:Demo}, we 
demonstrate possible ways one could utilize the proposed model 
by systematically comparing commonly used finite element 
packages and solvers. In Section \ref{Sec:Studies}, 
we extend the model to simulate nonlinear hydrostatic 
ice sheet flow equations. In Section \ref{Sec:Studies2}, we 
run the nonlinear hydrostatic ice sheet flow equations
across multiple compute nodes and study the performance. 
Concluding remarks and possible extensions of this work are 
outlined in Section \ref{Sec:Conclusions}.
All the notational conventions employed in this paper
are introduced as needed.

\section{COMMON PERFORMANCE ISSUES}
\label{Sec:Issues}
The performance of any scientific software or algorithm
will depend on a myriad of factors. First and foremost,
good performance depends on efficient and practical 
implementation of the code. Application and domain scientists 
may not be interested in the intricate details of the 
code framework that they did not design, but they must 
still be cognizant of important computational issues that 
may inhibit performance dramatically. 
We now briefly highlight some common performances issues 
computational scientist may come across in their line of work:
\begin{figure}[t]
\centering
\subfloat{\includegraphics[scale=0.42]{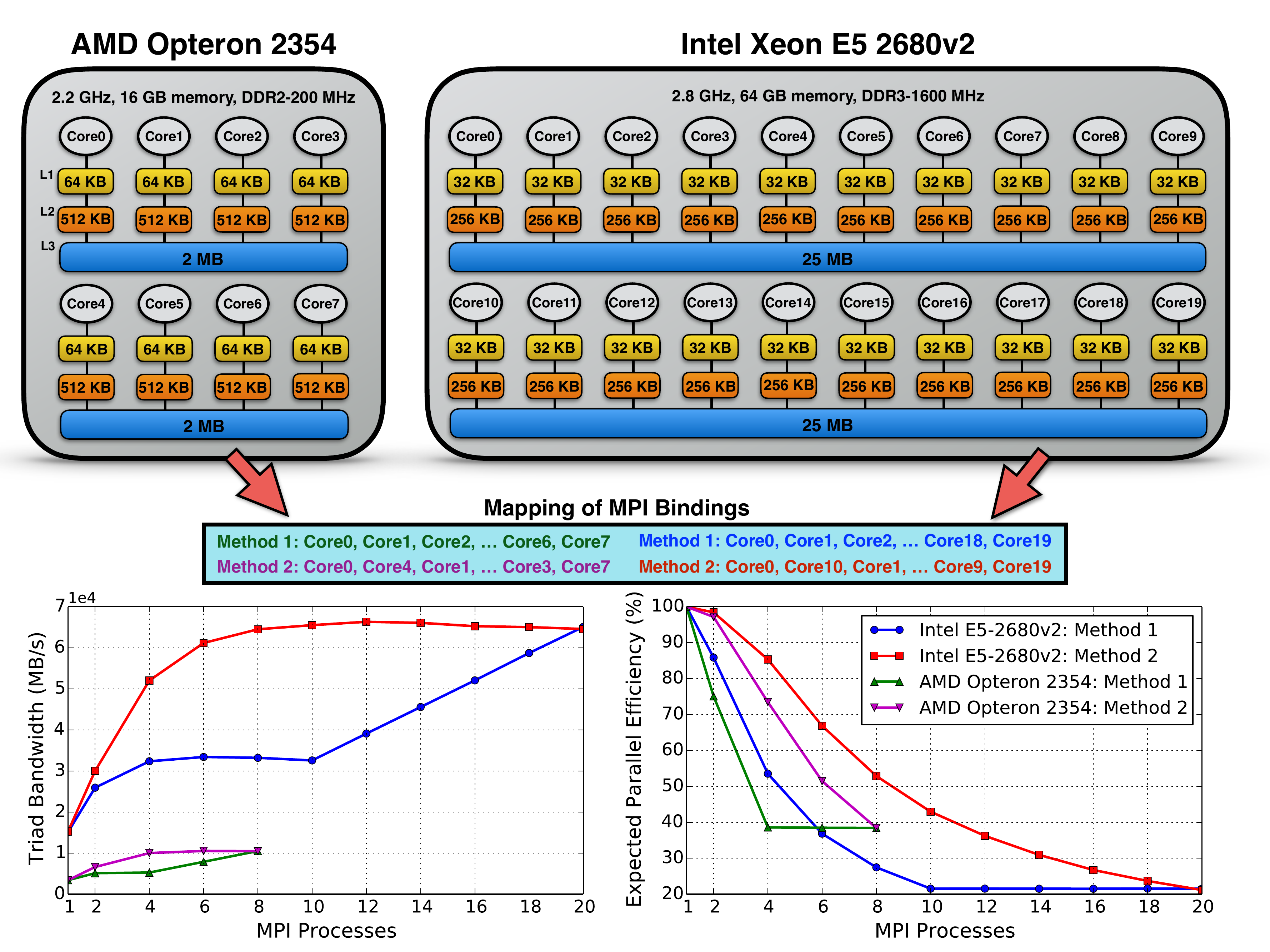}}
\captionsetup{format=hang}
\caption{An overview of the STREAM measurement on two different 
compute nodes. The mapping of MPI bindings has a significant impact on the 
achievable memory bandwidth.}
\label{Fig:streams_overview}
\end{figure}

\begin{itemize}
\item \textbf{Core/memory bindings}: The simplest way to maximize parallel
performance for MPI applications is to properly enforce MPI process and 
memory bindings. This is particularly 
important for memory bandwidth-limited applications because, on most CPU
architectures, the aggregate core bandwidth exceeds the CPU bandwidth to
memory and it is important to use the CPUs in a multi CPU server in a balanced way. 
Furthermore, if multiple users share a compute node, performance metrics 
can vary greatly as both memory resources and certain levels of cache are 
shared by others. Appropriate mapping 
methodologies for binding ranks to cores is vital for complex hardware architectures 
as well as for complex topological node layouts. Consider the single dual socket 
servers and their respective STREAM Triad benchmark results shown in Figure 
\ref{Fig:streams_overview}. Both the AMD and Intel processors possess two sockets 
where the physical cores are contiguously ordered. However, when 
the MPI processes are placed on alternating sockets, the achievable 
bandwidth is higher for a fixed number of cores by using the memory systems on 
both CPUs. For multi node performance, different binding techniques are 
required -- memory references on a single node are several times faster 
than on a remote node. Process allocation must be carefully done so 
that communication across networks is minimized.
\begin{table}
\centering
\captionsetup{format=hang}
\caption{Single node specifications from each of the HPC systems used for this study. 
Note that Intel's ``Knights Landing'' (KNL) processor has two different types of memory. \label{Tab:S2_HPC}}
{\tiny
\begin{tabular}{lccccc}
\hline
\multirow{3}{*}{\textbf{Processor}} & \textbf{AMD} & \textbf{Intel} & \textbf{Intel} & \textbf{Intel} & \textbf{Intel} \\
& \textbf{Opteron 2354} & \textbf{Xeon E5-2680v2} & \textbf{Xeon E5-2695v2} & \textbf{Xeon E5-2698v3} & \textbf{Xeon Phi 7250} \\
& \textbf{``Barcelona"} & \textbf{``Ivybridge"} & \textbf{``Ivybridge"} & \textbf{``Haswell"} & \textbf{``Knights Landing"} \\
\hline
Clock rate & 2.2 GHz & 2.8 GHz & 2.4 GHz & 2.3 GHz & 1.4 GHz\\
Year released & 2008 & 2013 & 2013 & 2014 & 2016 \\
Sockets & 2 & 2 & 2 & 2 & 1 \\
Cores/socket & 4 & 10 & 12 & 16 & 68\\
Threads/core & 1 & 1 & 2 & 2 & 4\\
\multirow{2}{*}{L1 cache} & 8$\times$64 KB & 20$\times$32 KB & 24$\times$64 KB & 32$\times$64 KB & 68$\times$64 KB \\
& 2-way associativity & 8-way associativity & 8-way associativity & 8-way associativity & 8-way associativity\\
\multirow{2}{*}{L2 cache} & 8$\times$512 KB & 20$\times$256 KB & 24$\times$256 KB & 32$\times$256 KB & 34$\times$1 MB\\
& 16-way associativity & 8-way associativity & 8-way associativity & 8-way associativity & 16-way associativity\\
\multirow{2}{*}{L3 cache} & 2$\times$2 MB & 2$\times$25 MB & 2$\times$30 MB & 2$\times$40 MB & - \\
& 32-way associativity & 20-way associativity & 20-way associativity & 20-way associativity & -\\
\multirow{2}{*}{Memory type} & \multirow{2}{*}{DDR2-200 MHz} & \multirow{2}{*}{DDR3-1600 MHz} & \multirow{2}{*}{DDR3-1866 MHz} & \multirow{2}{*}{DDR4-2133 MHz} & DDR4-2400 MHz, \\
& & & & & MCDRAM \\
\multirow{2}{*}{Total memory} & \multirow{2}{*}{16 GB} & \multirow{2}{*}{64 GB} & \multirow{2}{*}{64 GB} & \multirow{2}{*}{128 GB} & 96 GB (DDR4), \\
& & & & & 16 GB (MCDRAM) \\
\multirow{2}{*}{Memory channels} & \multirow{2}{*}{4} & \multirow{2}{*}{8} & \multirow{2}{*}{8} & \multirow{2}{*}{8} & 6 (DDR4), \\
& & & & & 8 (MCDRAM) \\
Compiler used & GNU & GNU & Cray & Cray & Cray \\
\multirow{2}{*}{STREAM Triad} & \multirow{2}{*}{10.5 GB/s} & \multirow{2}{*}{64.5 GB/s} & \multirow{2}{*}{102 GB/s} & \multirow{2}{*}{116 GB/s} & 90 GB/s (DDR4),\\
& & & & & 480 GB/s (MCDRAM) \\
\hline
\end{tabular}}
\end{table}
\item \textbf{Hardware architecture}: The performance of any benchmark or software 
depends on the hardware architecture. In this paper, we consider five different 
HPC systems with single node specifications listed in Table \ref{Tab:S2_HPC}.
It is evident from the STREAM Triad benchmark that 
different architectures have different levels of achievable memory-bandwidth.
Some of the processors are recent (as of the writing of this paper), 
like the Intel KNL processor whereas others like AMD's ``Barcelona" and 
Intel's ``Ivybridge" processors are older. With increasing complexity of processors
and memory systems, the challenge of good performance of solvers and 
algorithms has become an area of active research. A computational framework 
may solve a PDE efficiently on a laptop or small cluster, but that does not 
mean it will perform efficiently on a supercomputer. Understanding basic 
computer architectural concepts such as pipelining, instruction-level 
parallelism, and cache policies may offer excellent 
guidelines on how to speedup computations by several orders of magnitude. 
For example, Intel's KNL processor has two 512-bit vector units per core and may need
fine-grained parallelism to fully exploit the 68 cores per CPU. If a 
code is not properly vectorized to utilize the 136 vector units capable of 16
floating-point operations per cycle or the 16 GB of onboard
MCDRAM, it is possible that the speedup on this system will not be fully realized, 
and worse yet get outperformed by processors that have faster 
cores. Also, languages such as Python, which are used in some 
sophisticated finite element simulation packages, 
depend on the file system I/O because the interpreter executes system
calls to locate the module and may need to open hundreds of thousands of 
files before the actual computation can begin. Designing and utilizing 
algorithms/languages/compilers that are compatible with recent state-of-the-art 
HPC architectures is paramount \citep{Miller2013405}, otherwise the 
computational performance may be exceedingly poor. 
\begin{figure}[t]
\centering
\subfloat[Default ordering]{\includegraphics[scale=0.25]{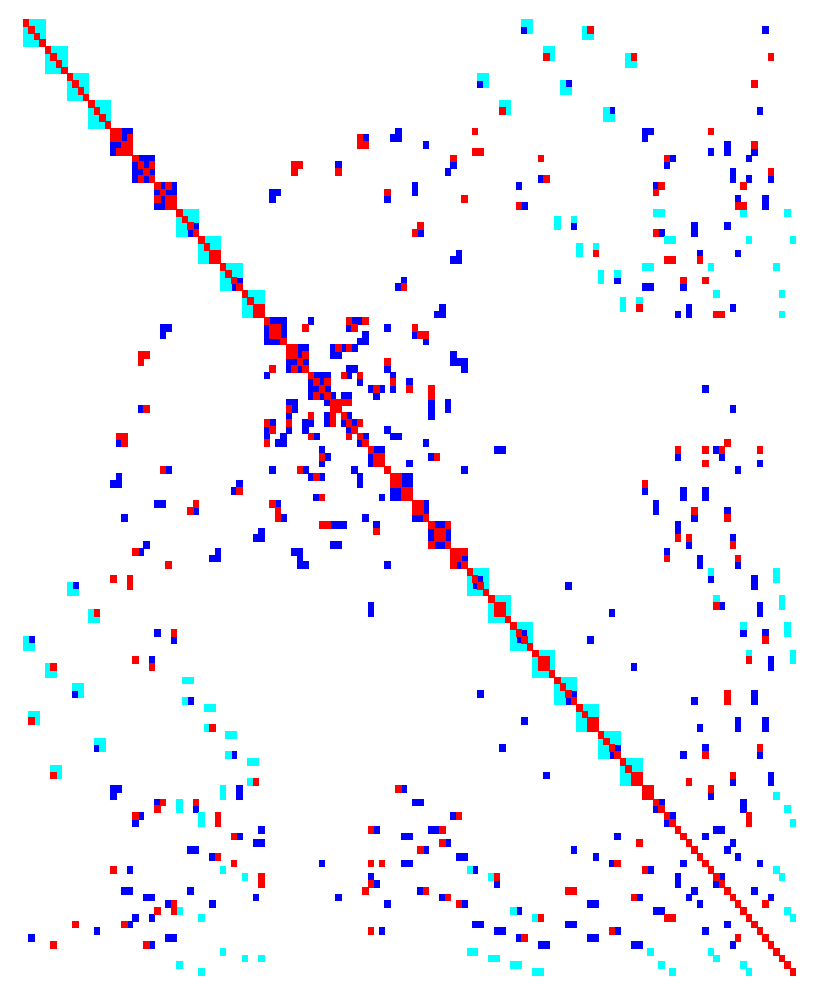}}
\subfloat[Optimized ordering]{\includegraphics[scale=0.25]{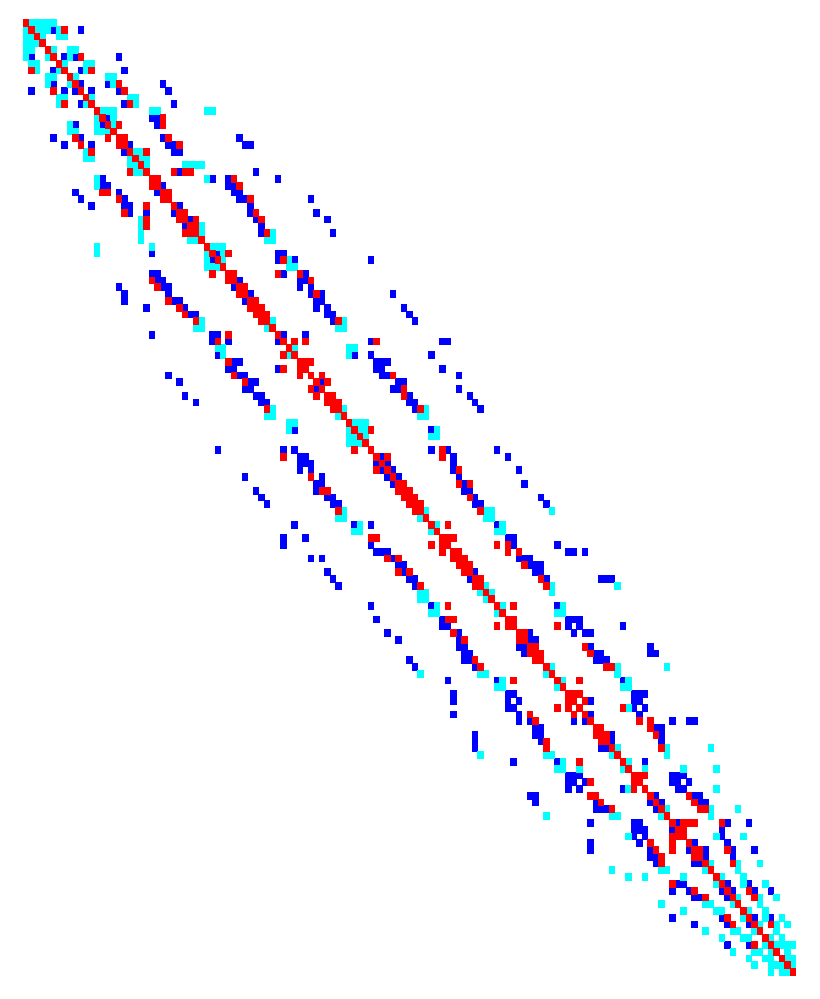}}
\captionsetup{format=hang}
\caption{Assembled sparse matrix where red represents positive 
numbers, blue represents negative numbers, and cyan represents allocated
but unused nonzero entries.}
\label{Fig:domain_decomposition}
\end{figure}
\item \textbf{Domain decomposition}: The global ordering and partitioning of the 
computational elements in a parallel computing environment, particularly for 
problems with unstructured grids, affect both spatial and 
temporal cache locality. Consider the assembled sparse matrices shown in
Figure \ref{Fig:domain_decomposition}. If the nonzero data entries are 
not properly grouped together, the code will invoke expensive cache misses 
and create little opportunity to use data in a cache line and reuse 
data in the cache. Consequently, this 
create serial bottlenecks at the cache/memory levels. Several mesh/graph 
partitioners such as Chaco \citep{hendrickson1995multi}, METIS/ParMETIS 
\citep{METIS}, and PTSCOTCH \citep{chevalier2008pt} are designed
to optimize locality and balance the workload among MPI processes. 
Some graph partitioners use a simple model of communication in seeking
to achieve load balance with minimum communication while others use a
more detailed communication model to better capture the minimum communication
needed by using hypergraphs instead of regular graphs as a basis for partitioning. 
Understanding which type of partitioning to use for the 
PDE problem at hand (e.g., spectral partitioning, geometric partitioning, 
multilevel graph partitioning, etc.) can significantly 
reduce the amount of communication and lead to higher efficiency 
and degree of concurrency.
\item \textbf{Solver convergence}: Arguably one of the most important
performance factors to consider for solving PDEs is the convergence rate 
of the solver. Direct methods like Gaussian elimination 
\citep{Grcar2011} as well as its sparse counterparts such as 
MUMPS \citep{amestoy2000mumps} and SuperLU\_DIST \citep{li2003superlu_dist} 
can solve problems in parallel but may have huge memory requirements as the problem
size is scaled up due to fill-in during factorization. Scalable and efficient solvers 
typically rely on the novel combination of iterative solvers and preconditioners.
The Krylov Subspace (KSP) and Scalable Nonlinear Equations Solvers (SNES) features
in the PETSc library coupled with robust preconditioners 
\citep{SmithBjorstadGropp1996,BrandtLivne2011} is a popular methodology
for solving large and complex PDEs. Novel combinations and tuning of solver 
parameters provide powerful and robust frameworks that can accurately 
and quickly converge to a specified residual tolerance, even for complex 
coupled multi-physics problems \citep{Castelletto2016894,bkmms2012,BruneKnepleySmithTu15}. 
Simple preconditioners 
such as Jacobi or Incomplete Lower Upper (ILU(0)) factorization may be 
fast for smaller problems, but the computational cost will soar because 
the number of solver iterations needed with Jacobi or ILU(0) will rapidly grow
with problem size. Scaling up the problem under these choices of preconditioning 
will be extremely time consuming and may not even converge for larger or more
complicated problems. Other more robust preconditioners, like the geometric 
and algebraic multigrid method, might have a more expensive setup time 
for smaller problems but have been demonstrated to maintain relatively 
uniform convergence for larger problems, even those that are nonsymmetric 
and indefinite \citep{bramble1994multigrid,Adams2004}.
\end{itemize}
These important performance issues should not be overlooked when 
analyzing the performance of a parallel computational framework. 
There are also several quick strategies for understanding and identifying 
bottlenecks on a specific HPC system. For example, it is well-known
that SpMV is an operation that is sensitive to the memory-bandwidth. These 
operations have very low AI's which can present itself as a bottleneck 
at the memory level on a single node. 
A simple test one can perform is to run the SpMV operation in parallel, and if 
it does not scale well on a single server in the strong sense, the memory-bandwidth 
is clearly limiting the performance. One can confirm 
this by running some simple vector operations like the vector sum and scalar multiplication to see if
they experience the same scaling issues. In addition, one can test the vector
dot product operation in order to detect problems with the
network interconnect or memory latency issues. The PETSc performance
summary~\citep{petsc-user-ref} provides comprehensive insight into the performance of many 
of these important operations including load balancing. The summary also 
provides information on the functions consuming most of the time. 
However, not all scientific software 
have readily available performance summaries, so a 
performance model amenable to any code implementation is needed to 
help answer common performance questions.

\section{PROPOSED PERFORMANCE SPECTRUM}
\label{Sec:Model}
\begin{figure}[t]
\centering
\subfloat{\includegraphics[scale=0.95]{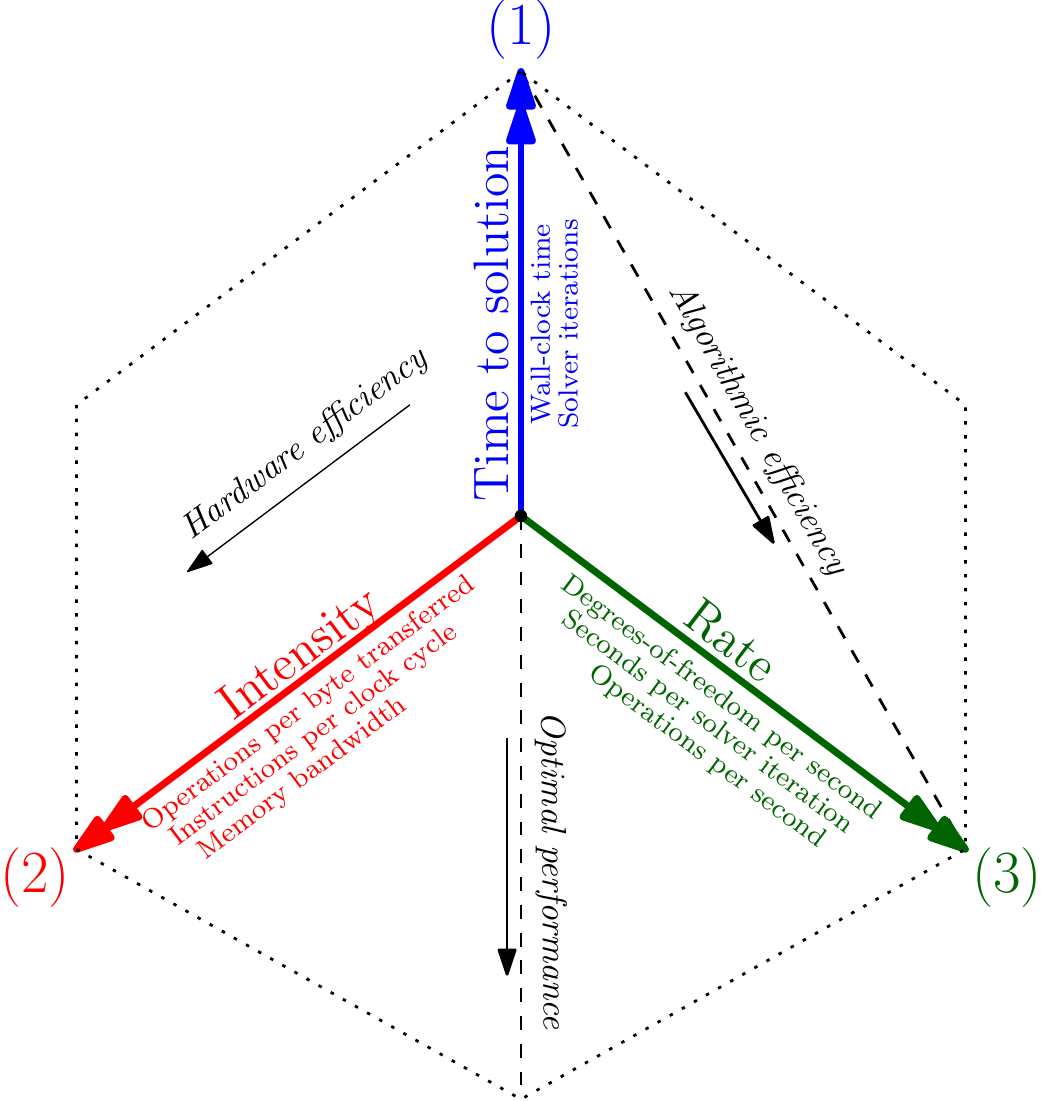}}
\captionsetup{format=hang}
\caption{Proposed
           performance spectrum that documents time, intensity and rate. 
           Intensity is defined as arithmetic intensity (FLOP to TBT ratio) based on 
           cache misses, and rate is defined as degrees-of-freedom solved per second.}
\label{Fig:performance_spectrum_outline}
\end{figure}

The general concept of the performance spectrum model is illustrated by
Figure \ref{Fig:performance_spectrum_outline}. This model
is designed to simultaneously capture both the hardware/architectural
exploitation as well as the algorithmic scalability of a particular 
parallel computational framework. First and foremost, we need the 
time-to-solution since this is the metric of most importance to 
application scientists needing to execute large-scale simulations on
state-of-the-art HPC systems. One may optionally document
the total number of solver iterations needed for convergence. However,
simply knowing the wall-clock time a computational framework needs 
to perform a task tells us little about the computational and algorithmic
efficiency. In order to understand how fast (or slow) a simulation is, we 
need to introduce two more metrics.

\subsection{Intensity}The second metric of interest is the intensity. 
Specifically, we focus on AI. As described in 
\citep{Williams_ACM_2009}, the AI of an algorithm or software is a 
measure that aids in estimating how efficiently the hardware resources and 
capabilities can be utilized. For the five machines listed in Table 
\ref{Tab:S2_HPC}, it is well-known that the limiting factor of performance 
for many applications is the memory-bandwidth. Thus, codes that have a high AI 
have a possibility of reusing data in cache and have lower memory bandwidth 
demands. It should be noted, however, that performance depends on many factors 
such as network latency and file system bandwidth, and the arithmetic intensity 
alone cannot be used to predict performance.

The general formula for the AI is defined as
\begin{align}
\label{Eqn:S3_AI}
\mbox{AI} := \frac{[\mbox{Work}]}{[\mbox{TBT}]},
\end{align}
where [Work] is the total amount of computational effort, typically what one
would refer to as FLOPs. The [TBT] metric is a measure of data movement between
the core/CPU and memory. A cache model is needed in order 
to not only determine the TBT but also to understand what amount of useful 
bandwidth is sustained for a given cache line transfer. One can employ any cache model for
this purpose, such as perfect cache, total number of load and store 
instructions at the core level, traffic at the memory level, or data 
cache misses. Different cache models are useful for interpreting different
behavioral trends, and the choice of cache model depends on the application or
research problem at hand. In this paper, we base [TBT] on the total number of
cache misses and cache line size. The formula for obtaining the TBT 
for the L1, L2, and L3 cache levels is expressed as
\begin{align}
\label{Eqn:S3_TBT}
\mbox{TBT}_{\mbox{Lx}} = \mbox{[Lx misses]}\times\mbox{[Lx line size (byte)]}.
\end{align}
The simplest way to define [Work] is as the total number of floating-point 
operations, denoted FLOPs. Thus the AI based on Lx cache misses is formally written as
\begin{align}
\label{Eqn:S3_AI_complete}
\mbox{AI}_{\mbox{Lx}} = \frac{[\mathrm{FLOPs}]}{\mbox{TBT}_{\mbox{Lx}}}.
\end{align}
If a solver or algorithm experiences a large number of cache misses at the last 
level, memory may impede performance. 

Sometimes the exact TBT of a particular algorithm is not of interest. Instead, 
an application scientist may only care about the relative measure, i.e., whether the 
AI is higher or lower compared to either another algorithm, a different implementation
of the same algorithm, or a different processor. Thus, one may simply look at the 
ratio of FLOPS and cache misses. Equation 
\eqref{Eqn:S3_AI_complete} may be simplified to
\begin{align}
\label{Eqn:S3_AI_simple}
\mbox{AI}_{\mbox{Lx}} = \frac{[\mathrm{FLOPs}]}{[\mbox{Lx misses}]}.
\end{align}
Every machine listed in Table \ref{Tab:S2_HPC} has a cache line size of 64
bytes for all levels of cache. Different CPUs may have different line sizes
and hence a cache miss may imply different memory demands on different
processor architectures. The remainder 
of the paper shall refer to the above formula for estimating the intensity
metric.
\begin{remark}
It should be noted that PAPI's methodology for counting FLOPS may be 
highly inaccurate for the ``Ivybridge'' systems listed in Table \ref{Tab:S2_HPC}.
The hardware counters only count the instructions issued and not the ones
executed or retired. This is paramount for iterative solvers
that rely on SpMV operations because as the codes spend time 
waiting for data to be available from memory, they will reissue the 
floating-point instructions multiple times. These reissues, coupled with 
incomplete filling of a vector unit instruction, can lead to 
overcount factors of up to 10 times. For a more thorough discussion 
on the issue of overcounts, see \citep{weaver2013overcount} and 
the references within. PAPI's FLOP counters 
are disabled on the ``Haswell'' and ``KNL'' processors due to the aforementioned issues
\end{remark}
\begin{remark}The AI can also be counted using other capable software tools and 
methodologies. For example, Intel's Software Development Emulator (SDE) can be used to obtain
FLOP counts \citep{intelSDE,flopSDE}  and Intel\textregistered VTune\texttrademark can be
used to obtain the TBT. This methodology is used in the commonly used Roofline model. 
Alternatively, one can approximate the FLOP count of 
a particular code by inserting counting mechanisms into the code. 
PETSc provides an interface and guidelines for manual FLOP counting, 
and thus FLOP counts for computational frameworks using it can be obtained 
through the performance summary output.
\end{remark}
\begin{remark}The correlation between AI and speedup on a single node 
may not always hold true in a cluster sense (i.e., scaling 
when communication networks are involved). The mechanisms used 
for MPI process info exchanged is very different when the 
processes are on the same node as opposed to on different nodes. 
An application scientist must be fully cognizant of not only
the HPC processor specification but also the network topology as well 
as the interconnect bandwidth and latency.
\end{remark}
\subsection{Rate}
Although $\mbox{AI}_{\mbox{Lx}}$ is useful for comparatively 
estimating the performance a particular parallel framework may attain, 
it does not necessarily aid in predictions of time-to-solution. Consequently, 
this means that AI can easily be ``gamed'' to appear high but the 
code consumes large amounts of wall-clock time. For example, 
small computationally intensive routines such as 
DGEMM \citep{Dongarra} can be inserted to artificially inflate the 
computational demands and increase the AI. Other performance models,
such as the Roofline model, would indicate that this is a favorable trend
while ignoring the fact that more time is spent than necessary. 
This is also why the traditional FLOP rate metric, which can also easily
be gamed, is not helpful either. 
Instead of measuring the total FLOPS executed per second, we measure the total 
degrees-of-freedom solved per second, hence the \emph{Rate} metric needed
to complete the performance spectrum is defined as
\begin{align}
\label{Eqn:S3_rate}
\mbox{Rate}_{1} := \frac{[\mathrm{DOFs}]}{[\mbox{total time (seconds)}]},
\end{align}
where $[\mathrm{DOFs}]$ simply refers to the total number of degrees-of-freedom or 
discrete component-wise equations that need to be solved.
\begin{definition}[Static-scaling]
Equation \eqref{Eqn:S3_rate} is an integral component of what we refer to as 
\emph{static-scaling}, where we increase the problem size but fix the 
concurrency. This is a complete reversal to the classical definition of 
strong-scaling where we fix the problem size but increase the concurrency. 
Static-scaling plots time-to-solution versus the total degrees-of-freedom solved
per second for a variety of problem sizes, so it also has characteristics similar
to the classical definition of weak-scaling where both problem size and concurrency
is increased. 
\end{definition}
\begin{figure}[t]
\centering
\subfloat{\includegraphics[scale=0.85]{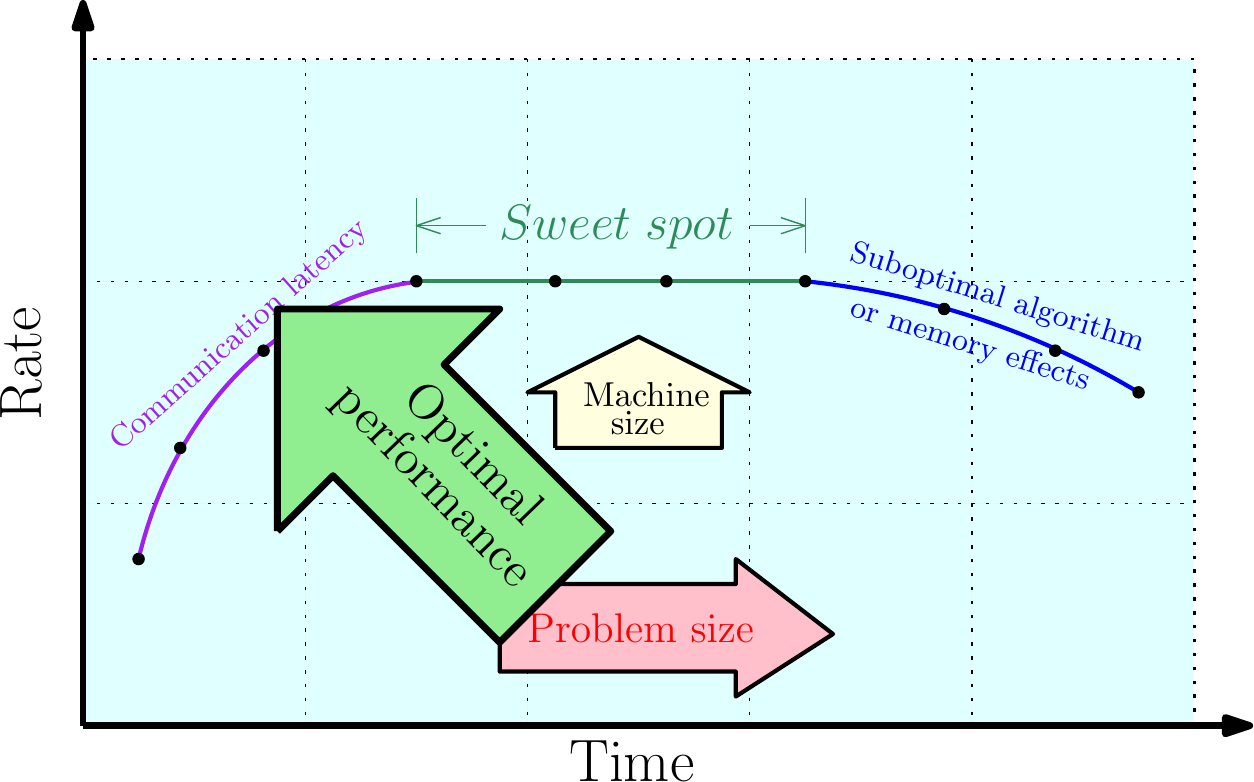}}
\captionsetup{format=hang}
\caption{Static-scaling plot. By fixing the MPI concurrency and increasing the 
problem size, the rate axis is the degrees-of-freedom solved per second.  
Algorithmic efficiency is achieved when a flat line is observed as the 
problem is scaled up.}
\label{Fig:algorithmic_goal}
\end{figure}

Figure \ref{Fig:algorithmic_goal} contains a pictorial description of a
static-scaling plot and illustrates how to visually interpret the data points. 
A scalable algorithm is $\mathcal{O}(n)$ where $n:=[\mathrm{DOFs}]$ is 
linearly proportional to $[\mbox{total time (seconds)}]$, so it is desirable to 
see a PDE solver maintain a constant rate metric for a wide range of 
problem sizes. The behavior of parallel computational frameworks for solving PDEs is 
not simple because 1) problems too small for a given MPI concurrency 
experience large communication to computation ratios (hence strong-scaling effects) 
and 2) large problems may have unfavorable memory accesses. The static-scaling plots are 
designed to capture both strong-scaling and weak-scaling characteristics and 
can give a good indicator of the ideal range of problem sizes for a given 
MPI concurrency.

The tailing off to the right of the static-scaling plot has two potential reasons.
First, problem size affects how memory is allocated and accessed. Larger problem
sizes may see an increase in memory contention as well as affect the access
pattern to main memory. Thus more time is spent waiting 
on data as opposed to performing calculations. However, another reason 
the tailing off occurs is because solvers for complex PDEs or computational domains may not always be 
$\mathcal{O}(n)$. Suboptimal algorithmic convergence may maintain a consistent
level of hardware utilization but require more iterations and 
FLOPs. To determine whether suboptimal algorithmic convergence plays a role 
in the deterioration of the static-scaling plot, equation \eqref{Eqn:S3_rate}
can be modified as
\begin{align}
\label{Eqn:S3_rate_iterate}
\mbox{Rate}_{2} := \frac{[\mathrm{DOFs}]}{[\mbox{time (seconds)}]\times
[\mbox{no. of solver iterations}]}.
\end{align}
This equation averages out increases in time due to an increase in iteration count.
If a flat line is observed using this metric, then poor algorithmic scalability
did have a negative impact on the static-scaling results.

Alternatively, if one is more interested in the performance gain for each MPI 
process, equation \eqref{Eqn:S3_rate} can also be modified into
\begin{align}
\label{Eqn:S3_rate_mpi}
\mbox{Rate}_{3} := \frac{[\mathrm{DOFs}]}{[\mbox{time (seconds)}]\times[\mbox{no. of MPI 
processes}]}.
\end{align}
This metric presents the average degrees-of-freedom solver per second for each MPI process
or core utilized. 
\subsection{Using the performance spectrum}
The arithmetic intensity and static-scaling components of the spectrum offer a variety of
strategies for interpreting the performance and scalability of any computational framework.
Good performance is achieved when a computational framework achieves low time-to-solution,
high arithmetic intensity, and flat static-scaling lines. The theoretical peak rate of
degrees-of-freedom solved per second could be unknown for a particular algorithm, but the
intensity metric can help us understand whether the static-scaling lines are good by estimating
how well it is efficiently using the available hardware resources. We outline three 
possible ways one could use the performance spectrum model:
\begin{enumerate}
\item \textsf{Hardware limitations:} As mentioned in Section \ref{Sec:Issues}, the
hardware configuration of the compute nodes plays a vital role in the performance spectrum
because different systems have different core counts, frequencies, and memory architectures.
Understanding how PDE solvers behave on different systems is vital for disseminating software 
to the scientific and HPC communities. The different cache sizes listed 
in Table \ref{Tab:S2_HPC} will be reflected in equation \eqref{Eqn:S3_rate}. 
AI$_{\mbox{Lx}}$ is likely to differ on different processors due to 
possible differences in cache sizes and cache policies. Furthermore, different
processors have different clock frequencies, arithmetic capabilities, and memory bandwidth.
Moreover, various GNU, Intel, and Cray compilers generate different executables that also
depend on optimization flags used. Compiled code also depend on the data structures 
used as well as code constructs. A particular platform may be better suited for certain 
PDE applications. The performance spectrum model is useful for quickly visualizing and 
comparing the impact of platform characteristics, software, compiler options, and algorithms.
\item \textsf{Software/solver implementation:} There are several software packages suited
for sophisticated finite element simulations such as the C++ based DEAL.II 
package \citep{BangerthHartmannKanschat2007}, the Python based Firedrake Project 
\citep{Rathgeber_ACM_2015}, the Python/C++ based FEniCS Project \citep{alnaes2015fenics},
the C++ based LibMesh \citep{libMeshpaper}, and MOOSE\citep{Gaston2009} projects.
These scientific libraries all use PETSc's linear algebra backend, but they can
also use other packages such as HYPRE \citep{hypre-users-manual} and 
Trilinos/ML \citep{trilinos:overview}. How well can specific solvers or software 
packages solve the same boundary value problem? Algebraic multigrid solvers
have various theoretical approaches and implementation strategies, so it is
entirely possible that certain solver configurations are better suited for 
a particular hardware architecture or PDE. Multigrid solvers for optimization 
remain a difficult research problem, but will be imperative for sustaining a 
high level of computational performance. 
Quick visual representations of the AI and equations solved per second
can certainly guide programmers and scientists in the right direction when designing
or implementing different software and solvers. 
\item \textsf{Numerical discretization:} Finally, various flavors of numerical 
discretizations such as the finite difference, finite element, and finite 
volume methods not only have different orders of mathematical accuracy
but different number of discrete equations to solve for a given mesh. Consider 
the Continuous Galerkin (CG) and Discontinuous Galerkin (DG) finite element methods -- clearly the 
DG method has more degrees-of-freedom since each element has its own copy of a geometric node,
but does that necessarily mean it is more time consuming? For example, if the CG and DG 
elements each take roughly $T$ seconds to attain a solution for the same computational
domain, then the latter element clearly has a higher rate metric because it
has more degrees-of-freedom for a given $h$-size, hence a bigger numerator in 
equation \ref{Eqn:S3_AI}. This is important for
computational scientists and mathematicians that want to compare the convergence rate
of various numerical methods particularly if $p$-refinement studies are involved. 
A cost benefit analysis can be performed when comparing the numerical accuracy vs
computational cost, often quantified using a work-precision diagram~\citep{KnepleyBardhan2015}.
One could also compare the impact finite element discretizations  
have on different geometric elements (e.g., tetrahedra, hexahedra, wedges, etc.). 
The performance of any numerical method depends on the hardware limitations and software
implementations, but this spectrum can be useful for comparing different
and available discretizations and polynomial orders in sophisticated finite element 
simulation packages.
\end{enumerate}

\section{DEMONSTRATION OF THE PERFORMANCE SPECTRUM}
\label{Sec:Demo}
As proof-of-concept, we apply the proposed performance spectrum 
to study the computational performance of a couple of popular finite 
element packages when used to solve the steady-state diffusion 
equation. A series of demonstrations shall enrich our current 
understanding of how hardware limitations, software implementation, 
numerical discretization, and material properties can impact the 
performance and scalability. We restrict our studies to the C++ 
implementation of the FEniCS Project and the Python implementation 
of the Firedrake Project, both of which leverage several 
scientific libraries and solvers such as PETSc, HYPRE, and Trilinos/ML 
solvers. The GMRES iterative solver is used with various 
algebraic multigrid solvers set to a relative convergence tolerance 
of $10^{-7}$. All numerical simulations
are performed on a single AMD Opteron 2354 and 
Intel Xeon E5-2680v2 node as described in 
Table \ref{Tab:S2_HPC}. In this section, the performance
spectrum model is used only to assess the assembly and solve steps.

The steady-diffusion equation gives rise to a second-order 
elliptic partial differential equation. To this end, 
let $\Omega$ denote the computational domain, and let 
$\partial \Omega$ denote its boundary. A spatial point 
is denoted by $\mathbf{x}$. The unit outward normal to 
the boundary is denoted by $\widehat{\mathbf{n}} (\mathbf{x})$. 
The boundary is divided into two parts: $\Gamma^{\mathrm{D}}$ 
and $\Gamma^{\mathrm{N}}$. The part of the boundary on which
Dirichlet boundary conditions are prescribed is denoted
by $\Gamma^{\mathrm{D}}$, and the part of the boundary on
which Neumann boundary conditions are prescribed is
denoted by $\Gamma^{\mathrm{N}}$. For mathematical well-posedness
we assume that
\begin{align}
  \Gamma^{\mathrm{D}} \cup \Gamma^{\mathrm{N}} = \partial \Omega
  \quad \mathrm{and} \quad
  \Gamma^{\mathrm{D}} \cap \Gamma^{\mathrm{N}} = \emptyset.
\end{align}
The corresponding boundary value problem
takes the following form
\begin{subequations}
  \label{Eqn:Steady_diffusion}
  \begin{alignat}{2}
    -&\mathrm{div}
    [\mathbf{D}(\mathbf{x})\mathrm{grad}[c(\mathbf{x})]] = f(\mathbf{x})
    &&\quad \mathrm{in} \; \Omega, \\
    &c(\mathbf{x}) = c^{\mathrm{p}}(\mathbf{x})
    &&\quad \mathrm{on} \; \Gamma^{\mathrm{D}},\quad\mathrm{and} \\
    -&\widehat{\mathbf{n}}(\mathbf{x}) \cdot \mathbf{D}(\mathbf{x})
    \mathrm{grad}[c(\mathbf{x})] = q^{\mathrm{p}}(\mathbf{x})
    &&\quad \mathrm{on} \; \Gamma^{\mathrm{N}},
  \end{alignat}
\end{subequations}
\begin{figure}[t]
\centering
\subfloat[Analytical solution]{\includegraphics[scale=0.42]{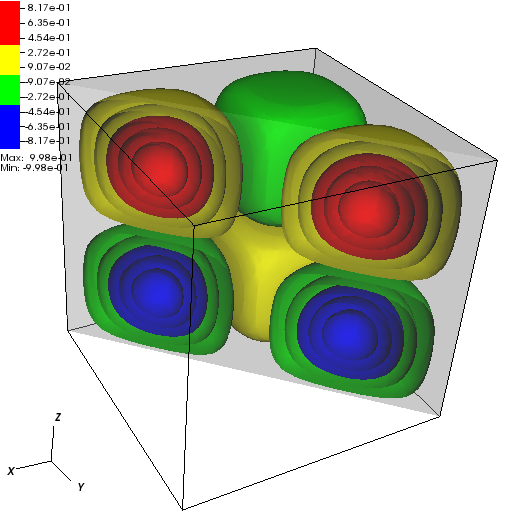}}
\subfloat[Mesh]{\includegraphics[scale=0.42]{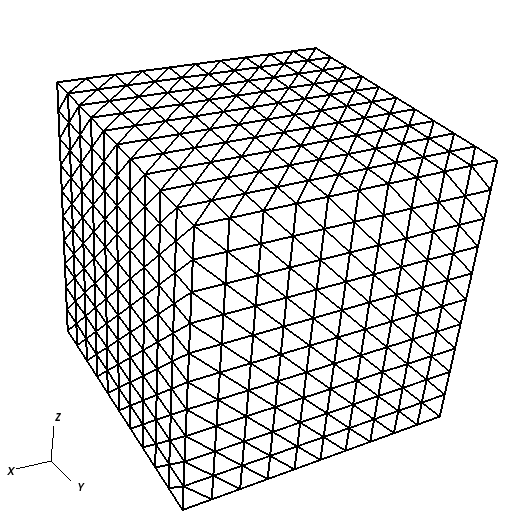}}
\captionsetup{format=hang}
\caption{Analytical solution of the steady-state diffusion example and the corresponding
mesh skeleton of the structure grid containing tetrahedra.}
\label{Fig:demo_solutions}
\end{figure}
\begin{table}
  \centering
\captionsetup{format=hang}
  \caption{Mesh discretization and CG1 $L_2$ error norm with respect to 
  $h$-refinement.
  \label{Tab:mesh}}
  \begin{tabular}{lcccc}
    \hline
    $h$-size & tetrahedra & Vertices & FEniCS $L_2$ error & Firedrake $L_2$ error\\
    \hline
    1/20 & 48,000 & 9,261 & 1.48E-02 & 2.96E-02  \\
    1/40 & 384,000 & 68,921 & 3.90E-03 & 7.77E-03 \\
    1/60 & 1,296,000 & 226,981 & 1.75E-03 & 3.51E-03 \\
    1/80 & 3,072,000 & 531,441 & 9.89E-04 & 1.99E-03 \\
    1/100 & 6,000,000 & 1,010,301 & 6.34E-04 & 1.28E-03 \\
    1/120 & 10,368,000 & 1,771,561 & 4.41E-04 & 8.88E-04 \\
    1/140 & 16,464,000 & 2,803,221 & 3.24E-04 & 6.52E-04 \\
    \hline
    & & & slope: 1.97 & slope: 1.96
  \end{tabular}
\end{table}
where $c(\mathbf{x})$ is the scalar concentration field, $\mathbf{D}(\mathbf{x})$ 
is the diffusivity coefficient, $f(\mathbf{x})$ is the volumetric source, 
$c^{\mathrm{p}}(\mathbf{x})$ is the prescribed concentration on the boundary, 
and $q^{\mathrm{p}}(\mathbf{x})$ is the prescribed flux on the boundary. 
Assuming $\mathbf{D}(\mathbf{x}) = \mathbf{I}$, we consider the 
following analytical solution and corresponding forcing function on a unit cube:
\begin{align}
  \label{Eqn:Analytical}
  &c(\mathbf{x}) = \mathrm{sin}(2\pi x)\mathrm{sin}(2\pi y)\mathrm{sin}(2\pi z)\quad\mathrm{and}\\
  &f(\mathbf{x}) = 12\pi^2\mathrm{sin}(2\pi x)\mathrm{sin}(2\pi y)\mathrm{sin}(2\pi z).
\end{align}
Homogeneous Dirichlet boundary conditions are applied on all faces, and the
analytical solution for $c(\mathbf{x})$ is presented in Figure 
\ref{Fig:demo_solutions}. These next few studies shall consider 
the following $h$-sizes on a structured tetrahedron mesh: 
1/20, 1/40, 1/60, 1/80, 1/100, 1/120, and 1/140. All mesh information 
and $L_2$ error norms with respect to the FEniCS and Firedrake 
implementations of the continuous Galerkin (CG1) element is 
listed in Table \ref{Tab:mesh}. 
\subsection{Demo \#1: AMD Opteron 2354 vs Intel Xeon E5-2680v2}
\begin{figure}[t]
\centering
\subfloat{\includegraphics[scale=0.5]{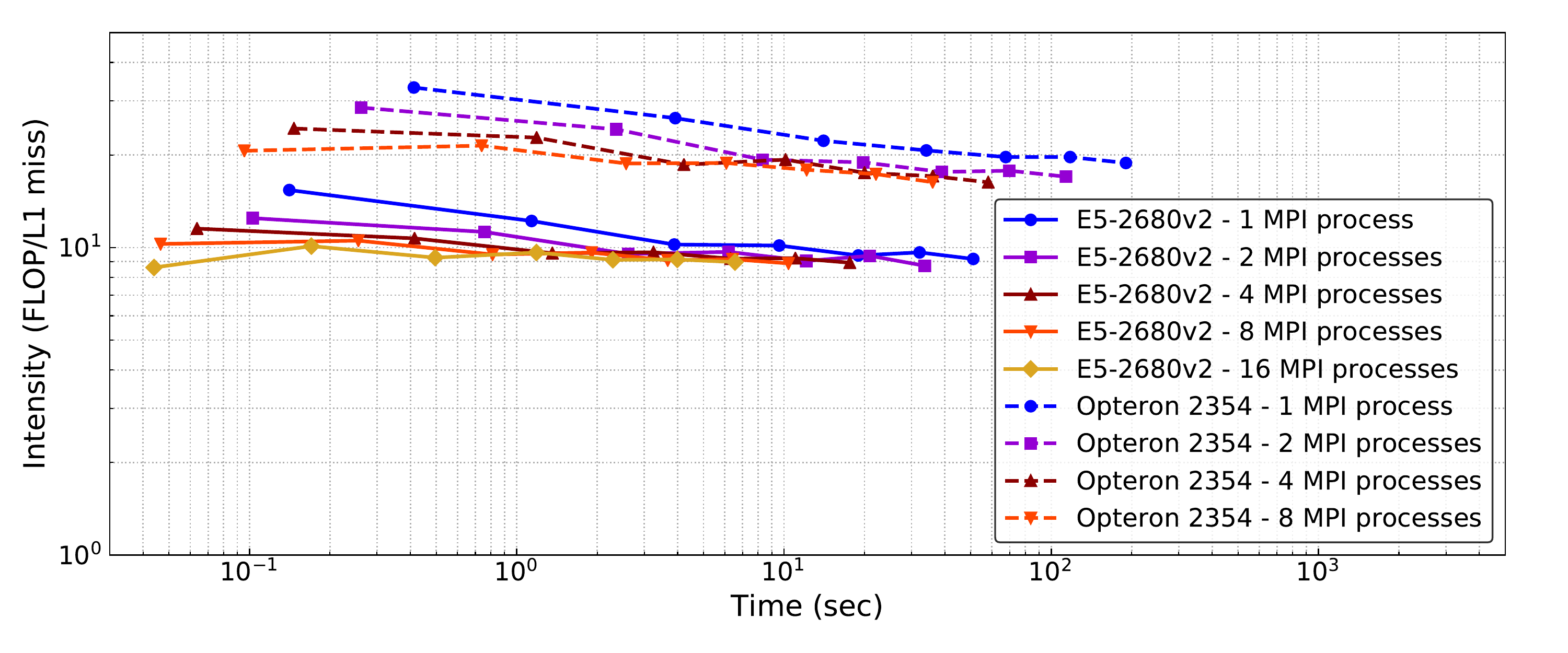}}
\captionsetup{format=hang}
\caption{Demo \#1: L1 arithmetic intensity for the FEniCS finite element package with 
PETSc's algebraic multigrid solver on a single Opteron 2354 and 
E5-2680v2 compute node.}
\label{Fig:demo1_intensity}
\end{figure}
\begin{figure}[t!]
\centering
\subfloat{\includegraphics[scale=0.5]{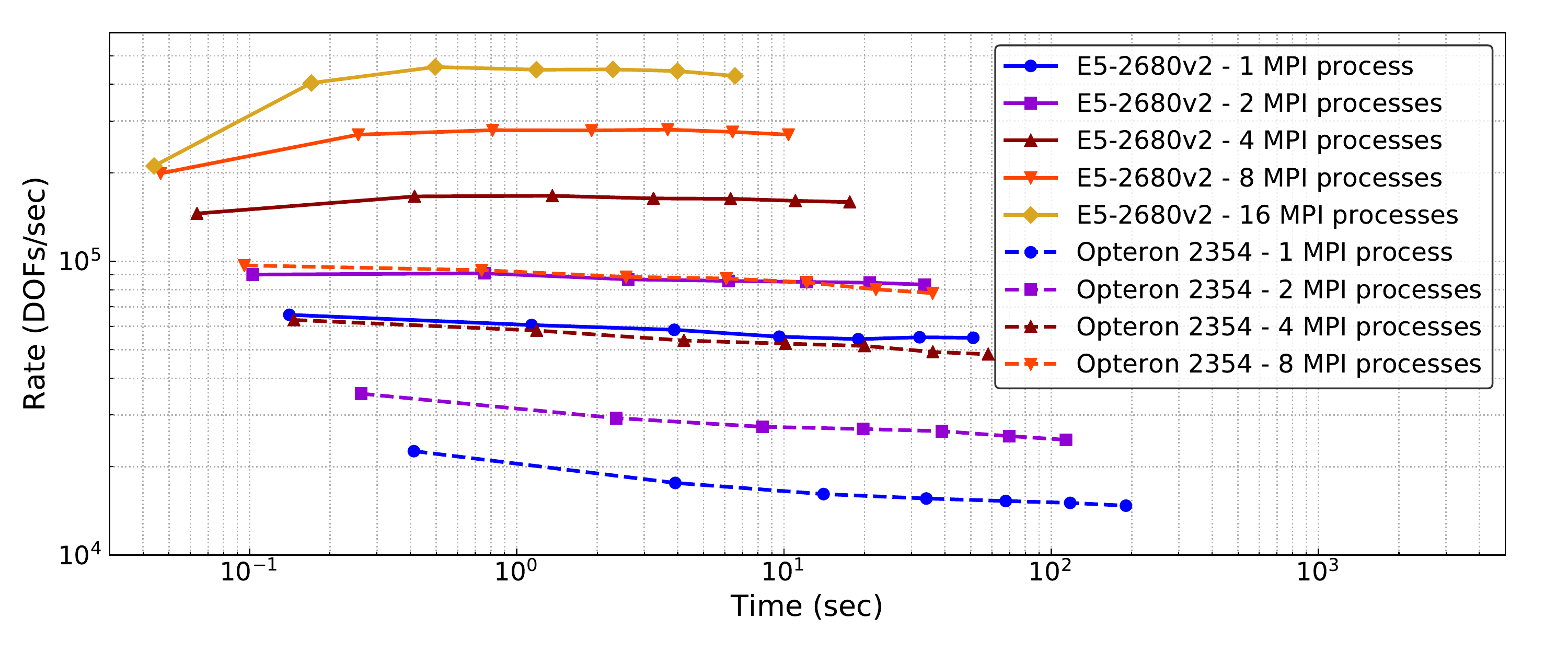}}
\captionsetup{format=hang}
\caption{Demo \#1: Static-scaling for the FEniCS finite element package with PETSc's 
algebraic multigrid solver on a single Opteron 2354 and 
E5-2680v2 compute node.}
\label{Fig:demo1_rate}
\end{figure}

\begin{figure}[t!]
\centering
\subfloat{\includegraphics[scale=0.5]{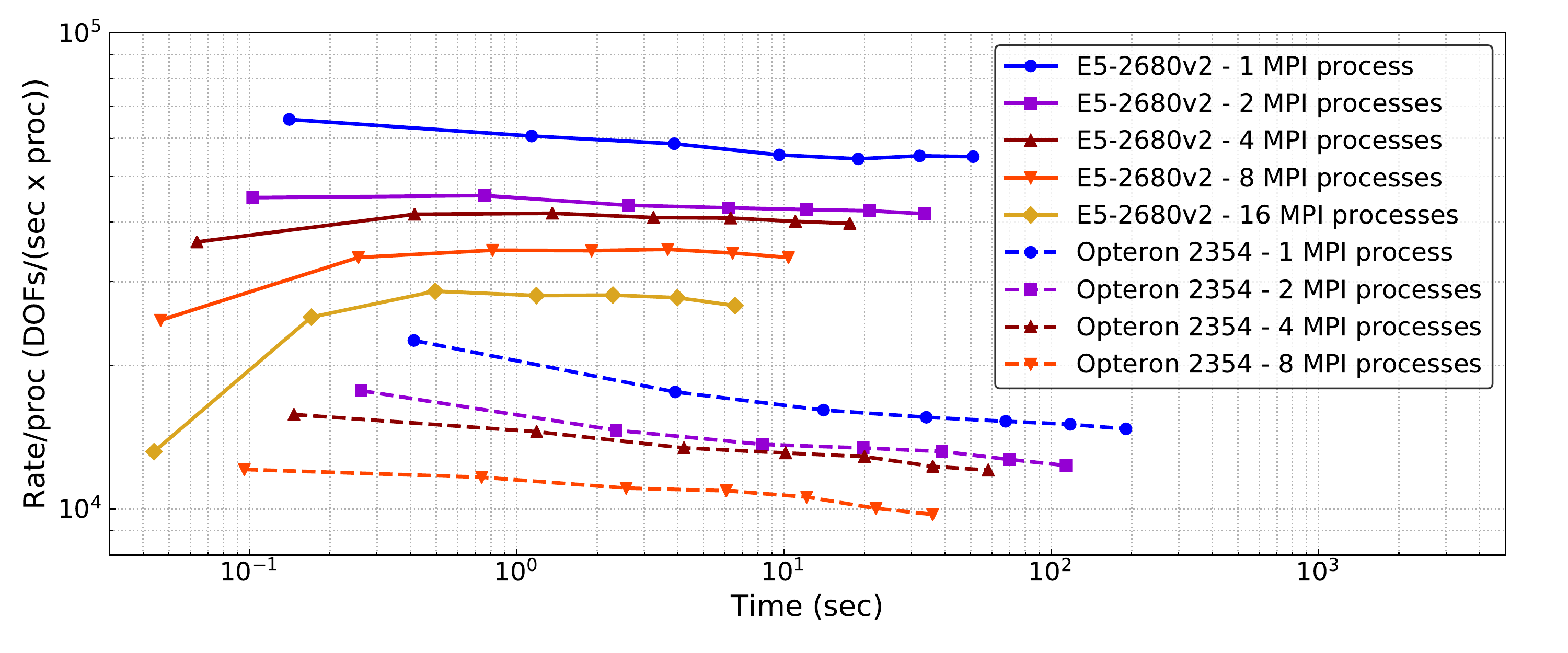}}
\captionsetup{format=hang}
\caption{Demo \#1: Static-scaling per MPI process for the FEniCS finite element package 
with PETSc's algebraic multigrid solver on a single Opteron 2354 and E5-2680v2 
compute node.}
\label{Fig:demo1_rateperproc}
\end{figure}

\begin{figure}[t]
\centering
\subfloat{\includegraphics[scale=0.5]{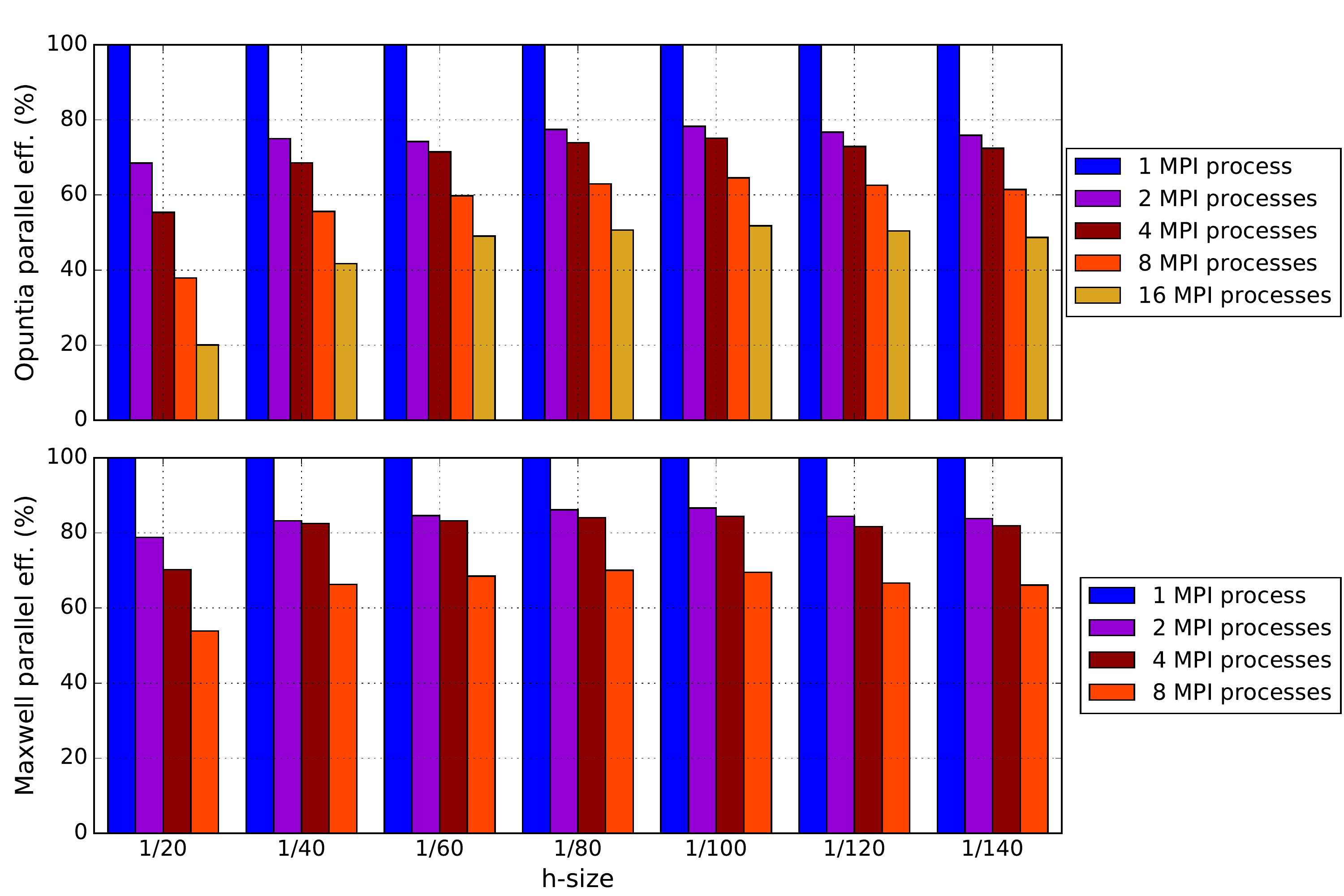}}
\captionsetup{format=hang}
\caption{Demo \#1: Strong-scaling efficiency for the FEniCS finite element package with PETSc's 
algebraic multigrid solver on a single Opteron 2354 and E5-2680v2 compute node.}
\label{Fig:demo1_speedup}
\end{figure}
We first compare the AI between a single Intel Xeon E5-2680v2 
and AMD Opteron 2354 compute node for FEniCS's implementation 
of the CG1 element coupled with PETSc's algebraic multigrid 
preconditioner. The AI$_{\mbox{L1}}$, 
as seen from Figure \ref{Fig:demo1_intensity}, gradually decreases with
mesh refinement. Moreover, increasing the 
number of MPI processes also reduces the AI. The Intel processor 
has smaller L1 and L2 caches compared to the AMD processor, 
which explains why the former processor has lower AIs. It can be 
concluded that a higher AI on a different machine does not 
necessarily translate to better performance because clock rates
and memory bandwidths differ. The fact that differences in 
AI$_{\mbox{L1}}$ does not directly relate to time-to-solution
can be seen in Figure \ref{Fig:demo1_intensity}.

The static-scaling plot is shown in Figure \ref{Fig:demo1_rate}. It is clear
that the Intel processor is capable of solving more degrees of freedom per 
second than the AMD processor. Increasing the number of MPI processes 
improves the Rate$_{1}$ metric, which is expected 
since time to solution is amortized. Employing Rate$_{3}$ from equation 
\eqref{Eqn:S3_rate_mpi}, as seen in Figure 
\ref{Fig:demo1_rateperproc} gives us a better insight into the effect adding
more MPI processes onto a single node has on the static-scaling performance. 
We also note that when only one or two MPI processes are used, 
the degrees-of-freedom solved per second metric degrades as the 
problem size increases. We also observe that the line plots for Intel reach 
higher apexes as more MPI processes and larger problems are solved. The 
lines curves ``dipping'' to the left indicate a degradation in parallel 
performance -- the problems are very small (e.g., $h$-size of 1/20 
resulting in 9,261 degrees-of-freedom distributed among 16 MPI 
processes means each process solves roughly only 580 equations) 
thus more of the execution time is spent on interprocess communication 
and latencies than actual computation. Both the Rate$_{1}$ and Rate$_{3}$ 
lines decrease with problem size on the AMD node, whereas the line plots 
for the Intel node are relatively flat, suggesting that the FEniCS and PETSc 
combination is in fact an algorithmically scalable combination for the 
problem at hand.

Figure \ref{Fig:demo1_speedup} depicts the parallel 
{efficiency} on the two different
nodes. The parallel performance for the smaller
$h$-sizes is significantly worse due to the lack of computation needed for 
a given MPI concurrency. It is interesting to note that the speedup 
on the AMD node is slightly greater than on the Intel node. 
Recalling from Figure \ref{Fig:demo1_intensity} that the AI$_{\mbox{L1}}$ 
on the AMD node is larger, we can infer that higher AIs indicate 
a stronger likelihood to experience greater parallel speedup. 
This behavior is consistent with the strong-scaling 
results of the optimization-based solvers for the Chromium 
remediation problem in \citep{Chang_JOMP_2016} where
similar classes of AMD and Intel processors were experimented with.
\begin{remark}
We note here that we expect speedup on a single node using 
MPI processes to be comparable to that achievable by threaded
implementations such as OpenMP or Intel TBB, since there are 
negligible system differences between threads and
processes~\citep{KnepleyBrownMcInnesSmithRuppAdams2015}. 
This kind of behavior has been observed in careful studies of
combined MPI+OpenMP finite element simulations~\citep{TurcksinKronbichlerBangerth2016}.
\end{remark}
\subsection{Demo \#2: FEniCS vs Firedrake}
\begin{figure}[t!]
\centering
\subfloat{\includegraphics[scale=0.5]{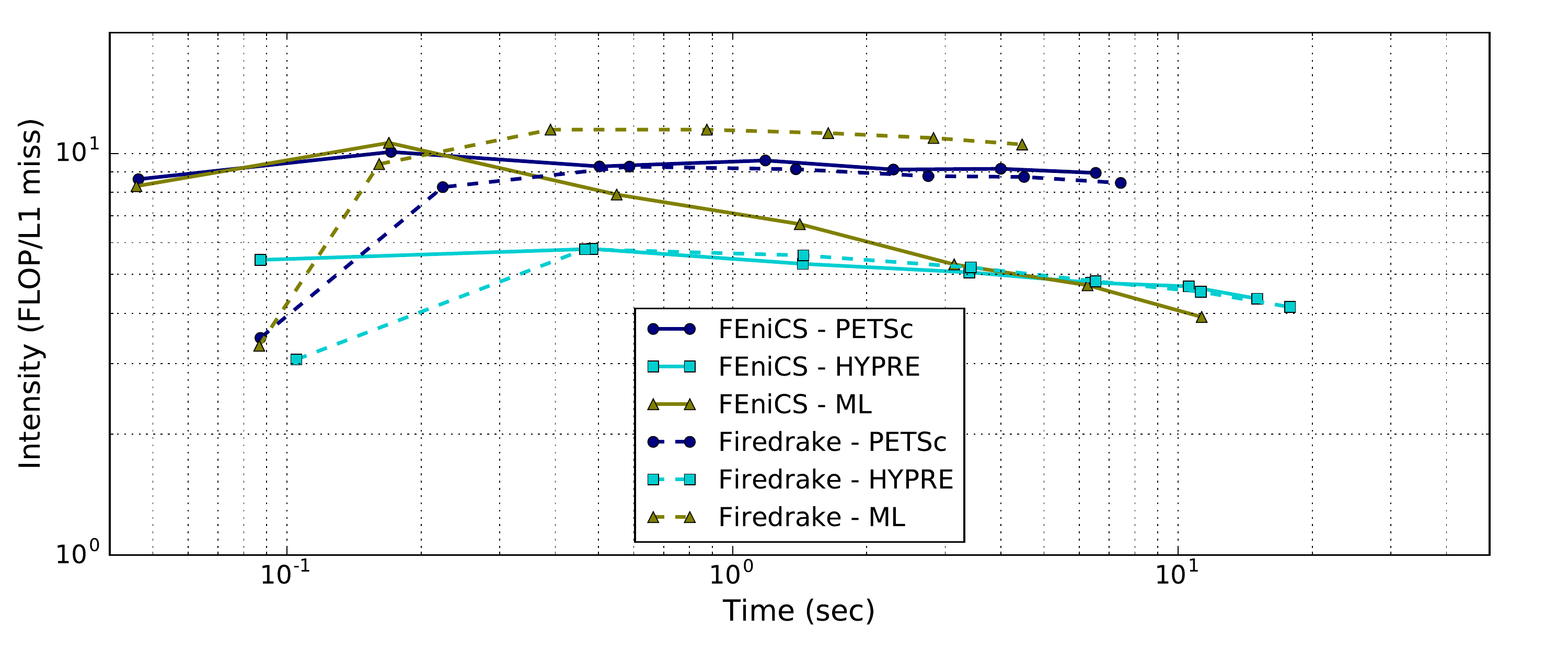}}
\captionsetup{format=hang}
\caption{Demo \#2: L1 arithmetic intensities for the FEniCS and Firedrake finite element packages 
with various solver packages on a single E5-2680v2 node with 16 MPI processes.}
\label{Fig:demo2_intensity}
\end{figure}

\begin{figure}[t!]
\centering
\subfloat{\includegraphics[scale=0.5]{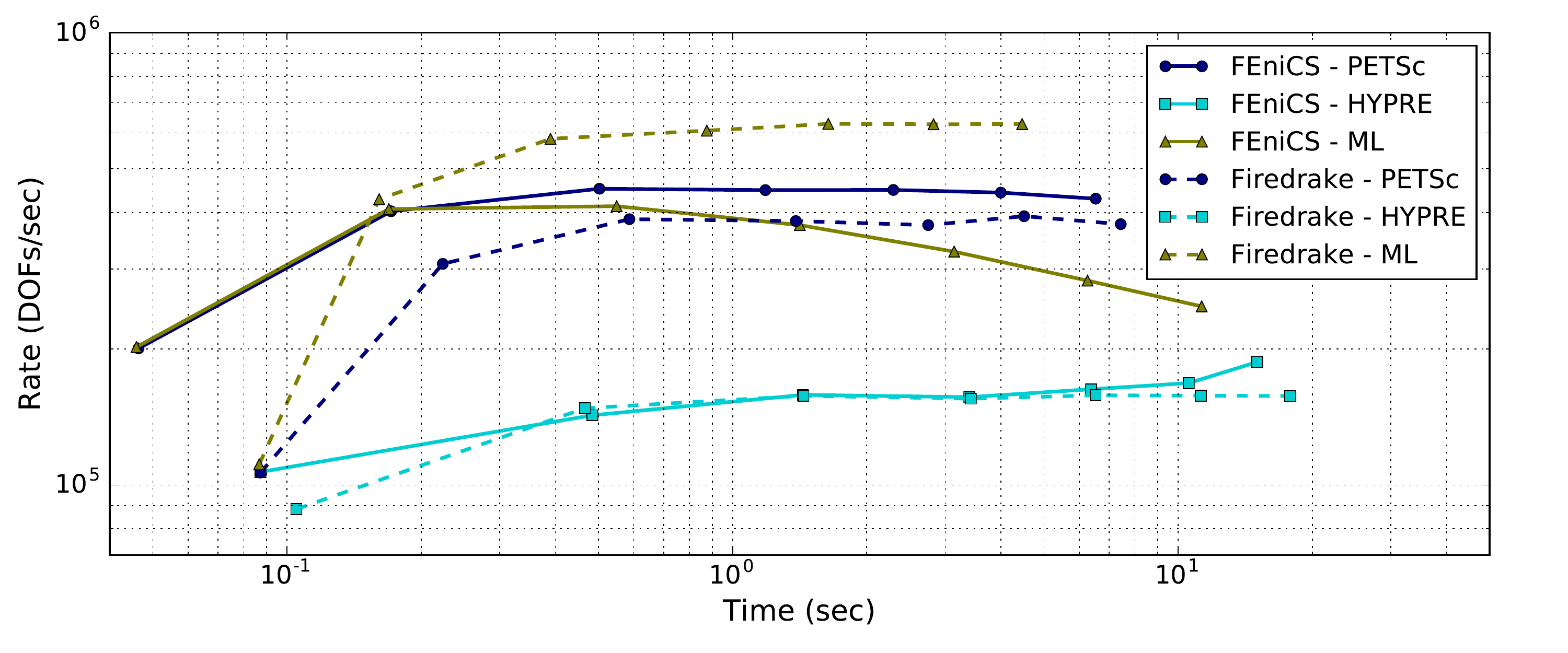}}
\captionsetup{format=hang}
\caption{Demo \#2: Static-scaling for the FEniCS and Firedrake finite element packages with 
various solver packages on a single E5-2680v2 node with 16 MPI processes.}
\label{Fig:demo2_rate}
\end{figure}

\begin{figure}[t!]
\centering
\subfloat{\includegraphics[scale=0.5]{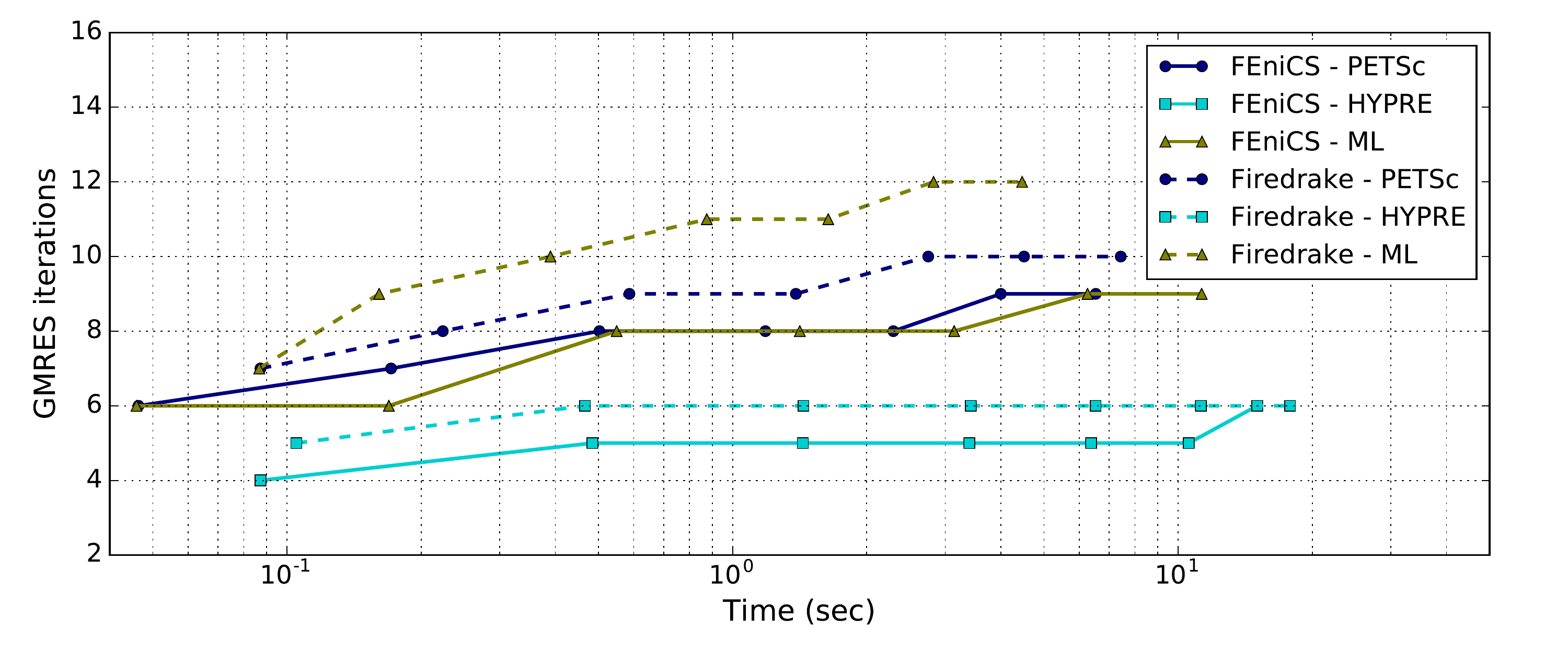}}
\captionsetup{format=hang}
\caption{Demo \#2: Number of GMRES iterations required for the FEnics and Firedrake finite element packages 
with various solver packages on a single E5-2680v2 node with 16 MPI processes.}
\label{Fig:demo2_iterations}
\end{figure}

\begin{figure}[t!]
\centering
\subfloat{\includegraphics[scale=0.5]{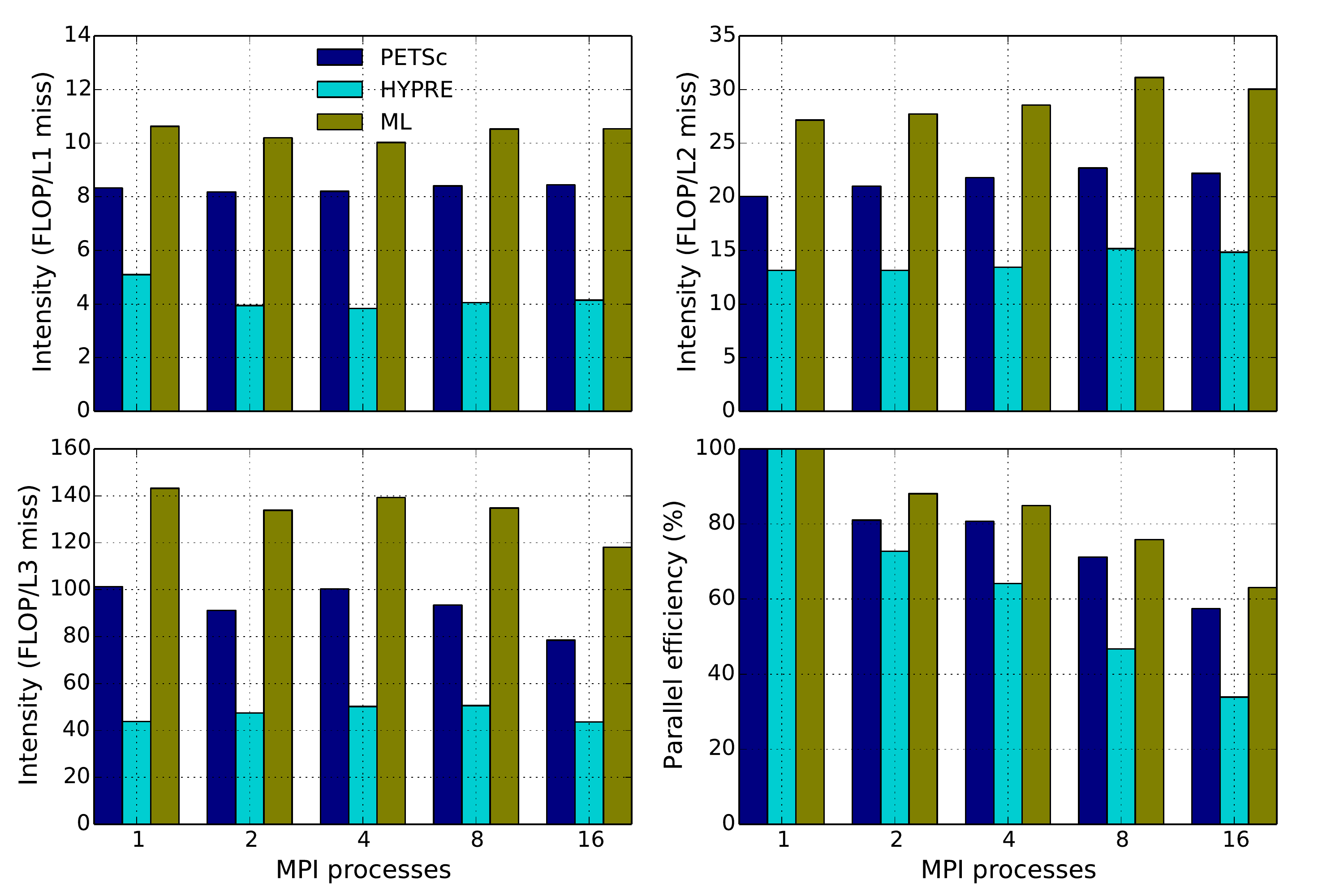}}
\captionsetup{format=hang}
\caption{Demo \#2: Correlation between the L1/L2/L3 arithmetic intensities and strong-scaling 
efficiency on a single E5-2680v2 node for up to
16 MPI processes when $h$-size = 1/140.}
\label{Fig:demo2_speedup}
\end{figure}

\begin{figure}[t]
\centering
\subfloat{\includegraphics[scale=0.5]{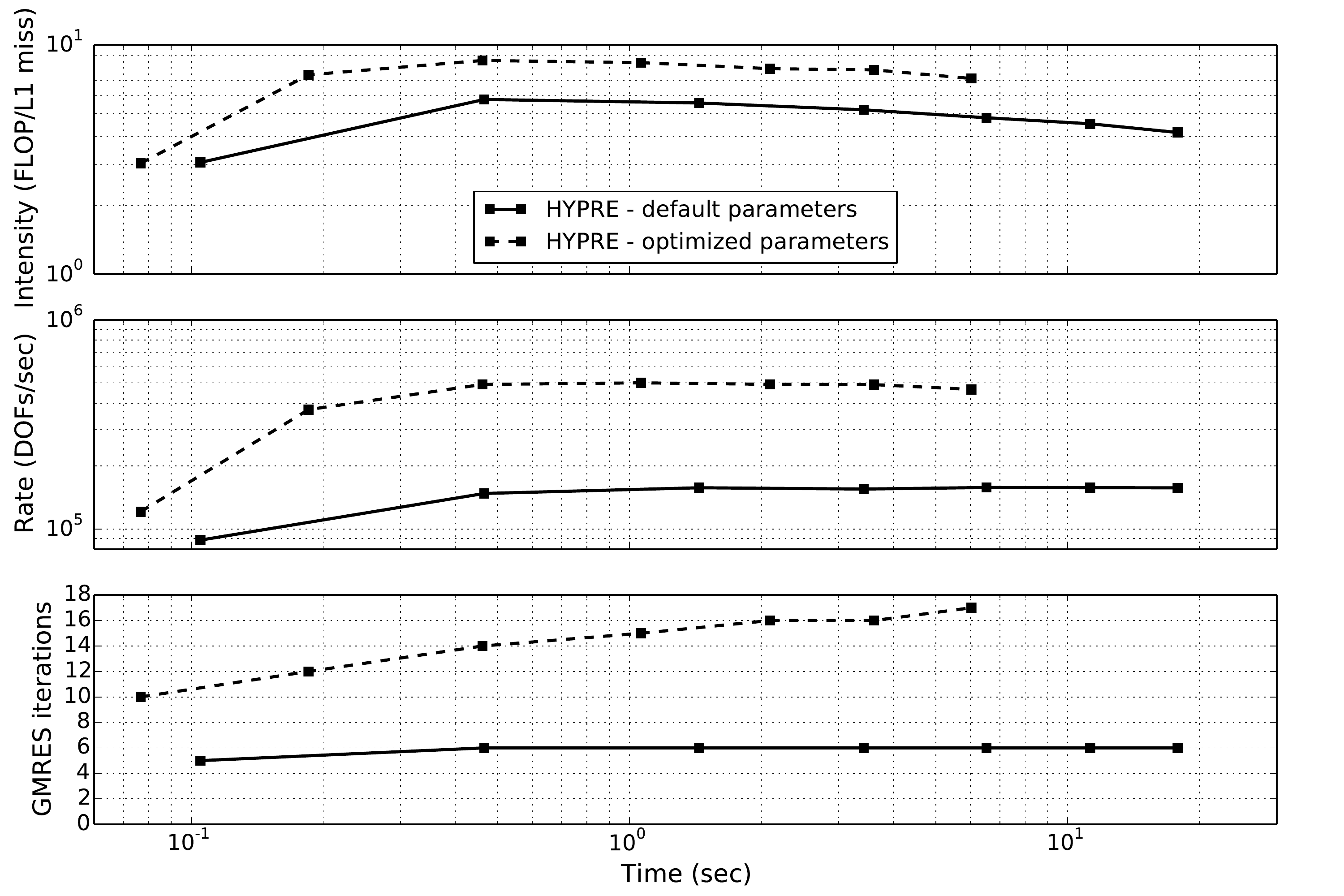}}
\captionsetup{format=hang}
\caption{Demo \#2: Comparison between HYPRE's default solver parameters and 
HYPRE's optimized solver parameters through the Firedrake package on a single 
E5-2680v2 node with 16 MPI processes.}
\label{Fig:demo2_hypre}
\end{figure}
Next, we compare the FEniCS and Firedrake implementations of the CG1 
element with 16 MPI processes on a single Intel Xeon E5-2680v2 node. The same 
steady-state diffusion equation is considered, but we now investigate 
how other multigrid solver packages like HYPRE and ML affect the 
performance.

The AI's in {Figure} \ref{Fig:demo2_intensity} clearly depend on 
the software implementation, the solver used, and the problem size. 
The results in this figure suggest that the FEniCS and
Firedrake packages have very similar implementations of the 
PETSc and HYPRE multigrid solvers. However, the AI$_{\mbox{L1}}$ 
for FEniCS's implementation of the ML solver deteriorates 
rapidly with problem size. Similar behavior is observed in 
the static-scaling plot of Figure \ref{Fig:demo2_rate} where 
the data points with the highest AI also have the highest 
rate at which equations are solved. Unlike the previous 
demonstration where different hardware implementations 
were compared, the AI and rate metrics are strongly correlated to each 
other, and it is clear that FEniCS's current implementation of 
ML has some issues since the tailing off towards the right 
occurs before either the PETSc or HYPRE lines do.

With these two graphs in mind, one may wonder why the tailing off occurs. 
Does it occur due to suboptimal algorithmic convergence 
(i.e., iteration count increases with problem size), or do the 
data structures needed for important solver steps begin to drop out of cache? A scalable 
algorithm suggests that the number of solver iterations should 
not increase by much when the problem size increases, so if 
the GMRES iteration count increases significant, it is possible that 
the rate metric will decrease. {Figure} \ref{Fig:demo2_iterations} 
denotes the number of GMRES iterations needed for every finite element package and 
solver, and it can be seen that the iteration counts do not increase by much. These 
plots must be interpreted carefully because although the iteration 
plots may suggest algorithmic scalability, the degradation in 
AI with respect to problem size suggests that the current software 
and solver parameters are not efficiently configured to utilize 
the hardware. As shown in the previous demonstration, 
the AI is useful for predicting which algorithms will see greater 
speedups as the number of MPI processes is increased. Figure 
\ref{Fig:demo2_speedup} compares the AI$_{\mbox{L1/2/3}}$ and 
parallel performance of Firedrake's three solver implementations. 
Regardless of which level of cache is used to determine the AI, 
HYPRE and ML have the lowest and highest AI's, respectively. Moreover, 
HYPRE and ML have the worst and best parallel speedups, respectively, which 
again supports the fact that the AI metric is useful for predicting which 
algorithms may achieve the greatest parallel speedup. 

We note that the HYPRE solver has relatively bad performance, 
suggesting that the out-of-box parameters are unfit for the problem at hand.
One of the best ways to improve the AI and Rate$_{1}$ metrics is to simply adjust 
some of the solver parameters. If, for example, we optimize the parameters
by increasing the strong threshold coarsening rate, the performance improves
dramatically as we can tell from Figure \ref{Fig:demo2_hypre}. The AI and 
Rate$_{1}$ metrics are now competitive with Firedrake's implementation of the 
PETSc and ML solvers, but it is important to realize that the GMRES 
iteration counts increased with size. An algorithm that requires fewer 
iterations yet remains constant when the problem size increase
does not necessarily mean it has good performance and scalability. 
Neither the AI nor rate metrics tail off towards the right, suggesting 
that the optimized HYPRE solver is scalable despite some minor growth 
in the GMRES iteration count. As we have discussed in the previous 
demonstration, answers regarding performance and scalability 
of various solvers and software will also depend on the hardware.

\subsection{Demo \#3: Continuous Galerkin vs Discontinuous Galerkin}
\begin{table}
  \centering
\captionsetup{format=hang}
  \caption{Demo \#3: Degrees-of-freedom with respect to $h$-refinement. 
  In this study we do not consider $h$-size = 1/100 for
  the DG1 or DG2 elements. \label{Tab:demo3_size}}
  \begin{tabular}{lcccc}
    \hline
    $h$-size & CG1 & CG2 & DG1 & DG2 \\
    \hline
    1/20 & 9,261 & 68,921 & 192,000 & 480,000 \\
    1/40 & 68,921 & 531,441 & 1,536,000 & 3,840,000 \\
    1/60 & 226,981 & 1,771,561 & 5,184,000 & 12,960,000 \\
    1/80 & 531,441 & 4,173,281 & 12,288,000 & 30,720,000 \\
    1/100 & 1,030,301 & 8,120,601 & - & -\\
    \hline
  \end{tabular}
\end{table}
\begin{table}
  \centering
\captionsetup{format=hang}
  \caption{Demo \#3: $L_2$ error norm with respect to $h$-refinement 
  for various finite elements provided through the Firedrake package.
  In this study we do not consider $h$-size = 1/100 for
  the DG1 or DG2 elements. \label{Tab:demo3_error}}
  \begin{tabular}{lcccc}
    \hline
    $h$-size & CG1 & CG2 & DG1 & DG2 \\
    \hline
    1/20 & 2.96E-02 & 3.81E-04 & 1.65E-02 & 2.16E-04 \\
    1/40 & 7.77E-03 & 3.79E-05 & 4.35E-03 & 2.26E-05 \\
    1/60 & 3.51E-03 & 1.06E-05 & 1.97E-03 & 6.47E-06 \\
    1/80 & 1.99E-03 & 4.44E-06 & 1.12E-03 & 2.72E-06 \\
    1/100 & 1.28E-03 & 2.25E-06 & - & -\\
    \hline
    slope: & 1.95 & 3.19 & 1.94 & 3.16
  \end{tabular}
\end{table}
\begin{figure}[t]
\centering
\subfloat{\includegraphics[scale=0.5]{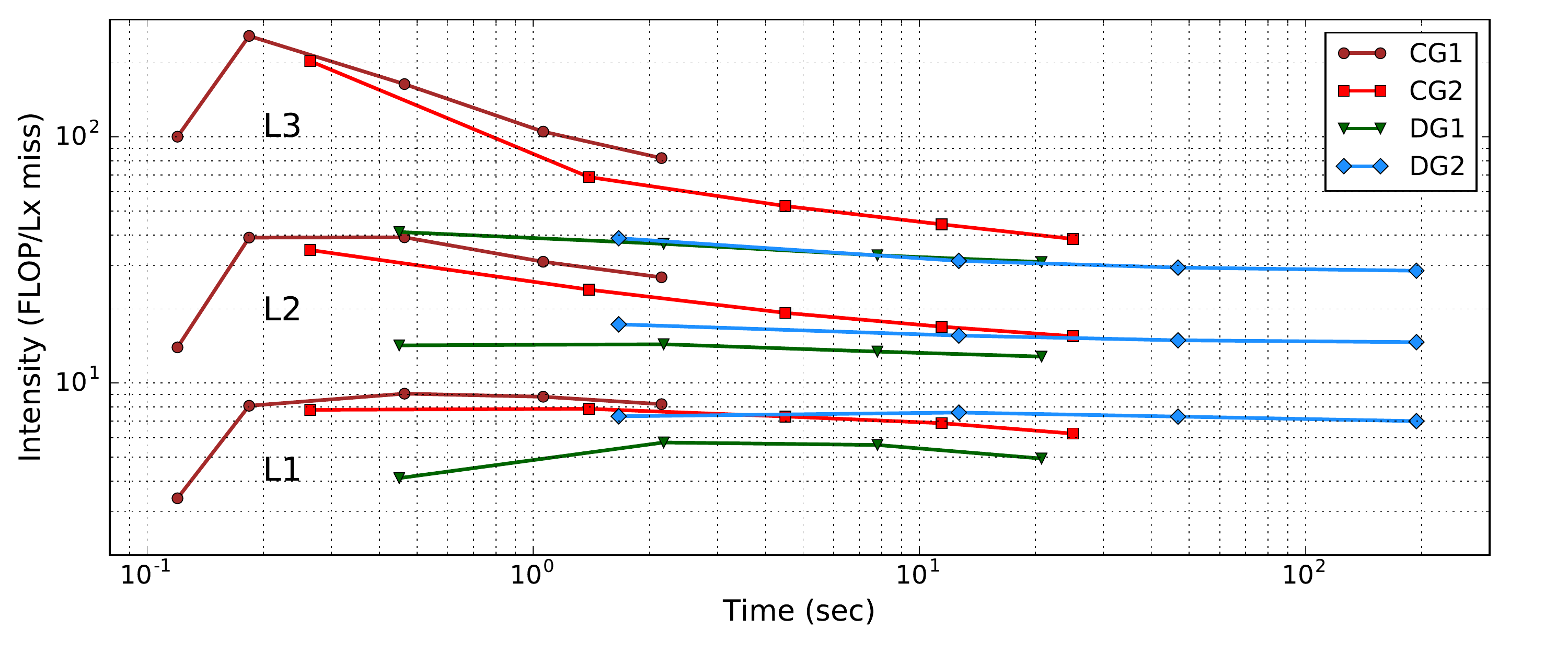}}
\captionsetup{format=hang}
\caption{Demo \#3: L1/L2/L3 arithmetic intensities of Firedrake's 
various finite element formulations on a single E5-2680v2 node
with 16 MPI processes.}
\label{Fig:demo3_intensities}
\end{figure}

\begin{figure}[t]
\centering
\subfloat{\includegraphics[scale=0.5]{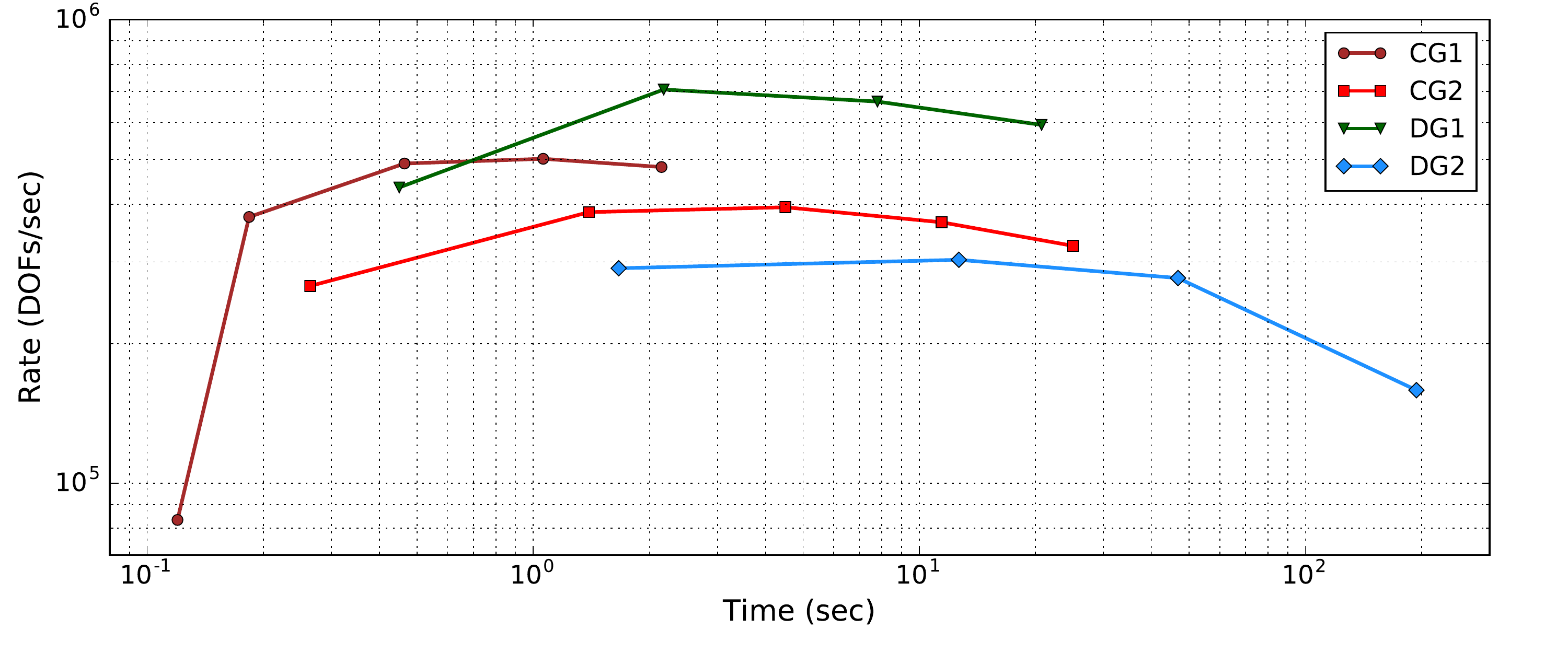}}
\captionsetup{format=hang}
\caption{Demo \#3: Static-scaling for Firedrake's 
various finite element formulations on a single E5-2680v2 node
with 16 MPI processes.}
\label{Fig:demo3_rate}
\end{figure}

\begin{figure}[t!]
\centering
\subfloat{\includegraphics[scale=0.5]{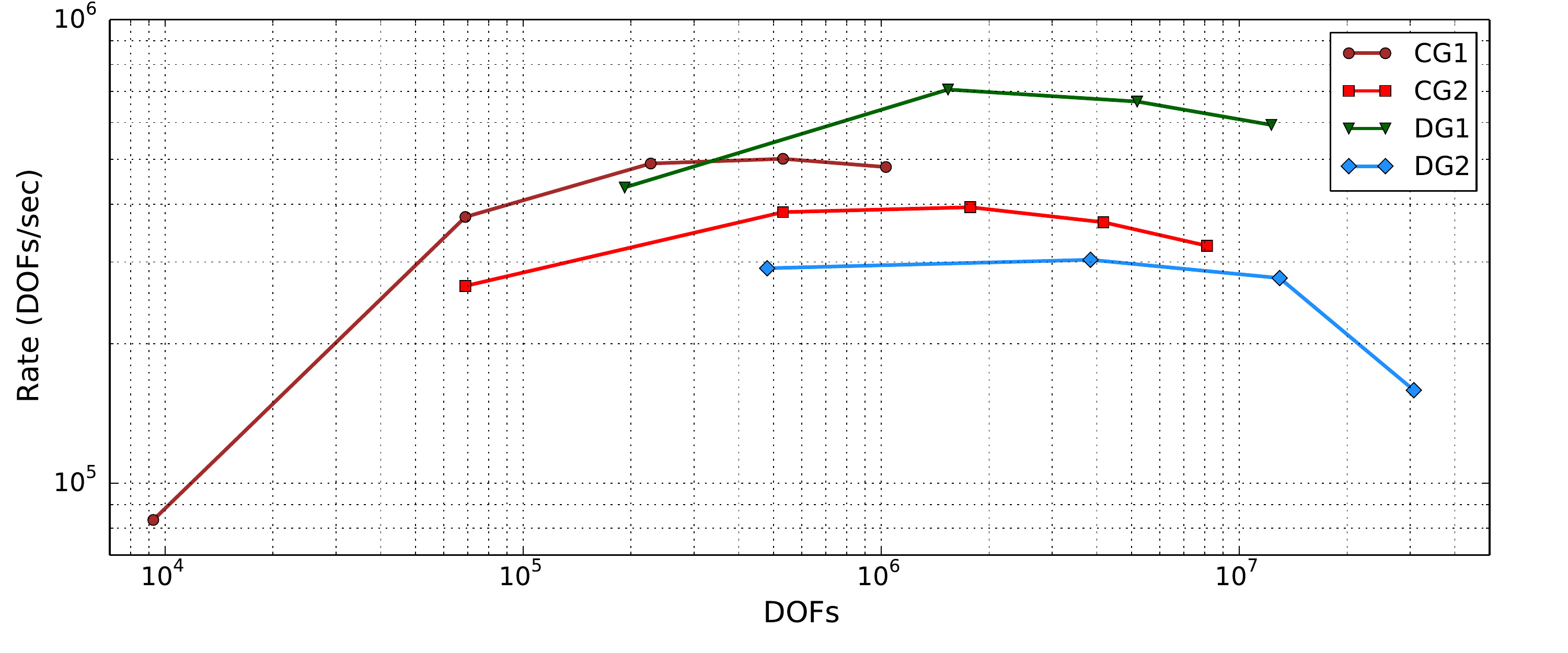}}
\captionsetup{format=hang}
\caption{Demo \#3: Degrees-of-freedom vs degrees-of-freedom solved per second 
for Firedrake's various finite element formulations on a single E5-2680v2 node
with 16 MPI processes.}
\label{Fig:demo3_rate3}
\end{figure}

\begin{figure}[t]
\centering
\subfloat{\includegraphics[scale=0.5]{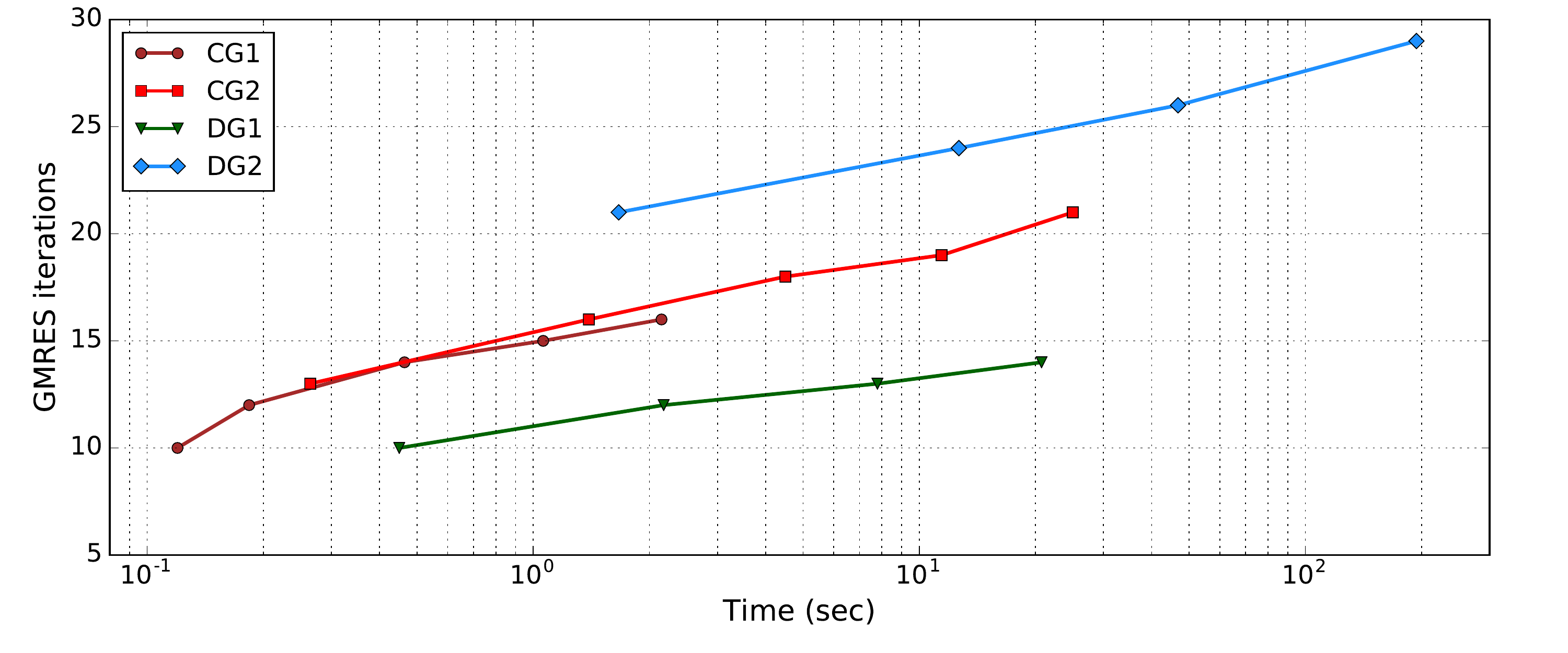}}
\captionsetup{format=hang}
\caption{Demo \#3: Solver iterations needed for Firedrake's 
various finite element formulations on a single E5-2680v2 node
with 16 MPI processes.}
\label{Fig:demo3_iterations}
\end{figure}

\begin{figure}[t!]
\centering
\subfloat{\includegraphics[scale=0.5]{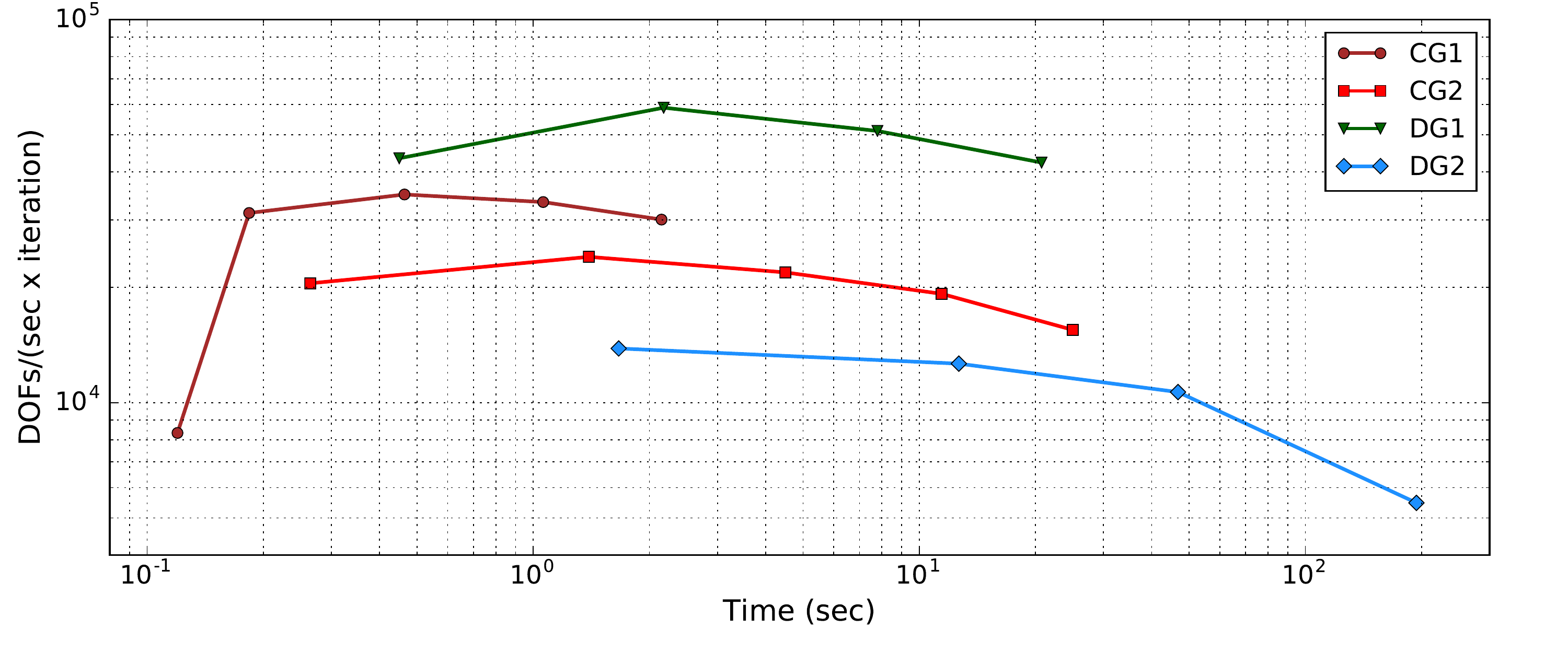}}
\captionsetup{format=hang}
\caption{Demo \#3: Static-scaling per solver iteration for 
Firedrake's various finite element formulations on a single 
E5-2680v2 node with 16 MPI processes.}
\label{Fig:demo3_rateperiterate}
\end{figure}
So far we have only considered the CG1 finite element. What happens if we
employ another discretization such as the Discontinuous Galerkin (DG) method? 
Moreover, what happens if we increase the polynomial order and employ second
order CG (CG2) and second order DG (DG2) elements? Various families of 
elements and their respective levels of $p$-refinement will change both 
the size and numerical accuracy of the numerical solution, so it is
desirable to understand both the costs and benefits of these approaches
on a particular mesh. Tables \ref{Tab:demo3_size} and 
\ref{Tab:demo3_error} contain the total degrees-of-freedom and $L_2$ error norms,
respectively, of Firedrake's various finite element discretizations. The
CG elements are studied up to $h$-size = 1/100 whereas the DG elements are
studied up to $h$-size = 1/80. We again employ 16 MPI processes across 
a single Intel Xeon E5-2680v2 node, and all finite element discretizations in this 
demonstration are solved with optimized (i.e., increased strong threshold 
coarsening) HYPRE parameters. 

Figure \ref{Fig:demo3_intensities} contains the AI$_{\mbox{L1/2/3}}$ for the
CG1, CG2, DG1, and DG2 elements. What we learn from these results is that 
increasing the polynomial order for the CG elements lowers the AI whereas 
the AI increases for DG elements. This may not always be the case because 
different solvers and different hardware architectures may be better 
tailored to different discretization. Other finite element packages like the
FEniCS or DEAL.II projects may have very different results. The Rate$_{1}$ metric 
as seen from Figure \ref{Fig:demo3_rate} depicts the rate at which each 
discretization solves its equations. Alternatively, one could also compare
the Rate$_{1}$ metric with respect to the degrees-of-freedom as seen in Figure
\ref{Fig:demo3_rate3}. Although DG elements have more degrees-of-freedom
for a given mesh discretization, it is seen that the DG1 element has the highest
Rate$_{1}$ metric, suggesting that the optimized HYPRE solver parameters are 
especially suitable for DG1 elements. Unlike the FEniCS and ML combination
example from the previous demonstration, the DG2 discretization experiences 
significant degradation in the static-scaling plot yet maintains relatively
consistent AI's. This begs the question of whether the tailing off towards
the right is due to memory effects or suboptimal algorithmic convergence.

As previously observed from Figure \ref{Fig:demo2_hypre}, 
the optimized HYPRE parameters resulted in a slight increase 
in GMRES iteration count for CG1 elements, and we notice similar
trends for the other finite elements in Figure \ref{Fig:demo3_iterations}. If the
iteration count increase is significant enough, it could negatively affect 
static-scaling. To determine whether this solver iteration growth stymied the rate 
by which equations are solved, we can employ Rate$_{2}$ (i.e., degrees-of-freedom solved per 
second per solver iterate) from equation \eqref{Eqn:S3_rate_iterate}
as shown in Figure \ref{Fig:demo3_rateperiterate}. In this particular demonstration, 
it makes no difference as we still observe degradation with respect to problem 
size, hence suggesting that memory bandwidth and cache behavior have 
an adverse effect on the simulation. Using more
compute nodes may certainly ameliorate both the AI and rate metrics for
the DG2 element, but it should again be cautioned that comparative studies 
on the performance of numerical methods and solvers strongly depend on 
both the code implementation as well as the nature of the computing platform.

\subsection{Demo \#4: Material properties}
\begin{table}
  \centering
\captionsetup{format=hang}
  \caption{Demo \#4: $L_2$ error norm with respect to $h$-refinement for various values
  of $\alpha$ in equation \eqref{Eqn:anisotropic_diffusivity} when using the 
  Firedrake implementation of the CG1 element.
  \label{Tab:demo4_error}}
  \begin{tabular}{lcccccc}
    \hline
    $h$-size & $\alpha = 0$ & $\alpha = 1$ & $\alpha = 10$ & $\alpha = 100$ & $\alpha = 1000$ \\
    \hline
    1/20 & 1.48E-02 & 3.45E-02 & 4.83E-02 & 5.71E-02 & 5.86E-02 \\
    1/40 & 3.90E-03 & 9.31E-03 & 1.46E-02 & 1.90E-02 & 1.99E-02 \\
    1/60 & 1.75E-03 & 4.23E-03 & 6.84E-03 & 9.26E-03 & 9.76E-03 \\
    1/80 & 9.89E-04 & 2.40E-03 & 3.94E-03 & 5.43E-03 & 5.75E-03 \\
    1/100 & 6.34E-04 & 1.55E-03 & 2.55E-03 & 3.55E-03 & 3.77E-03 \\
    1/120 & 4.41E-04 & 1.07E-03 & 1.78E-03 & 2.49E-03 & 2.64E-03 \\
    1/140 & 3.24E-04 & 7.88E-04 & 1.31E-03 & 1.84E-03 & 1.96E-03 \\
    \hline
    slope: & 1.97 & 1.94 & 1.86 & 1.77 & 1.75
  \end{tabular}
\end{table}
\begin{figure}[t]
\centering
\subfloat{\includegraphics[scale=0.5]{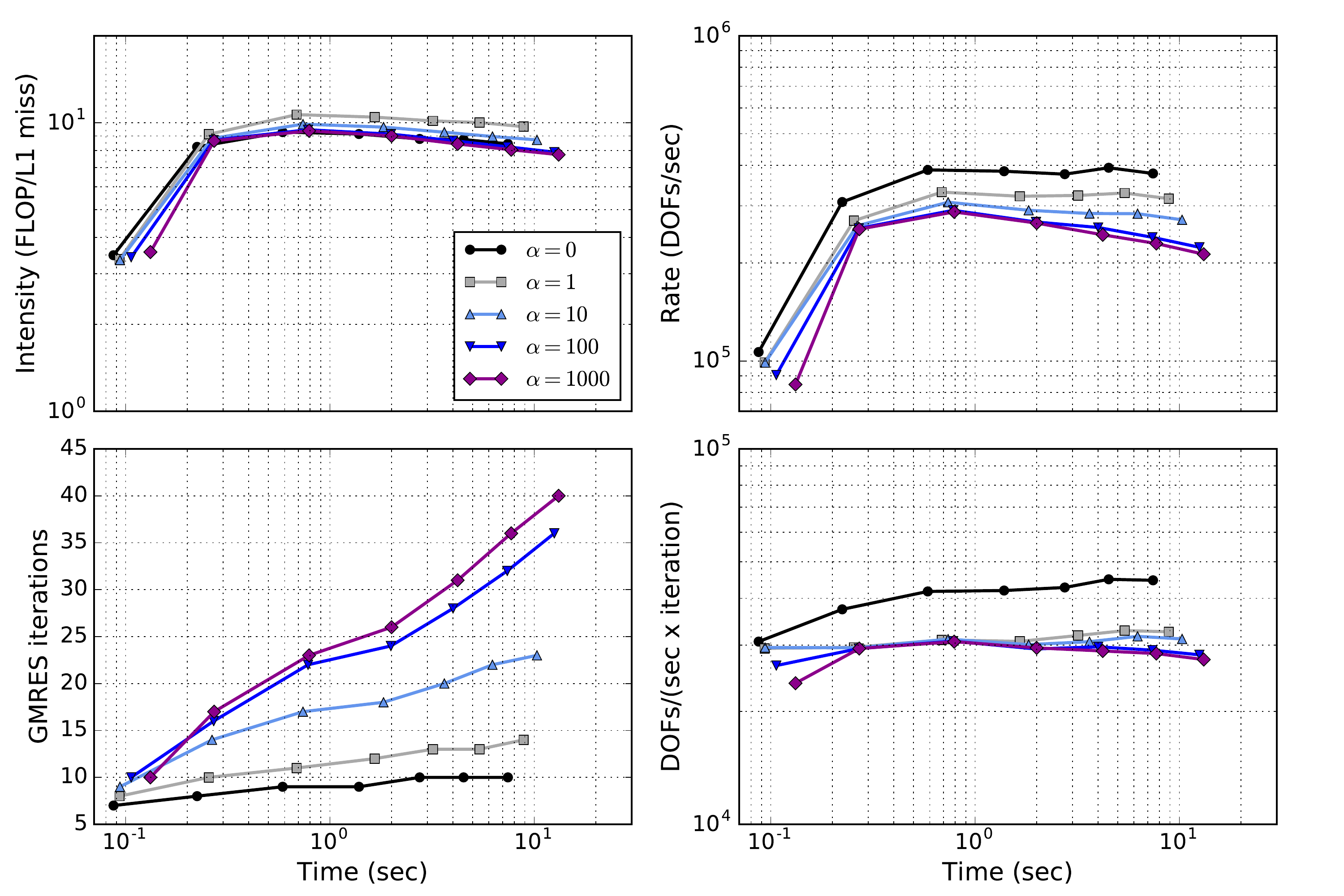}}
\captionsetup{format=hang}
\caption{Demo \#4: Performance spectrum for various values of $\alpha$ in equation \eqref{Eqn:anisotropic_diffusivity} 
on a single E5-2680v2 node with 16 MPI processes.}
\label{Fig:demo4_spectrum}
\end{figure}
So far all three of our demonstrations have been conducted in a homogeneous 
domain. However, many scientific problems are often heterogeneous in nature, 
which may complicate the physics of the governing equations and may become more 
expensive to solve numerically. In \citep{Chang_CMAME_2017}, it was shown that
solving heterogeneous problems like chaotic flow resulted in suboptimal 
algorithmic convergence (i.e., the iteration counts grew with $h$-refinement), 
so our goal is to demonstrate how physical properties such a heterogeneity 
and anisotropy may skew how we interpret the performance. 
Let us now assume that we have a heterogeneous 
and anisotropic diffusivity tensor that can be expressed as follows
\begin{align}
  \label{Eqn:anisotropic_diffusivity}
   &\mathbf{D}(\mathbf{x}) = \left(\begin{array}{ccc}
  \alpha(y^2+z^2)+1 & -\alpha xy & -\alpha xz \\
  -\alpha xy & \alpha(x^2+z^2)+1 & -\alpha yz \\
  -\alpha xz & -\alpha yz & \alpha(x^2+y^2)+1 
  \end{array}
  \right),
\end{align}
where $\alpha \geq 0$ is a user defined constant that controls the 
level of heterogeneity and anisotropy present in the computational domain. 
By employing the same analytical solution as equation \eqref{Eqn:Analytical}, 
the various values of $\alpha$ give rise to new forcing functions. The $L_2$ error 
norms with respect to $\alpha$ using Firedrake's CG1 elements are shown in Table
\ref{Tab:demo4_error}. Again a single Intel Xeon E5-2680v2 compute node with 
16 MPI processes is used for this study, and PETSc's multigrid solver is 
used to solve these problems.

Figure \ref{Fig:demo4_spectrum} depicts the AI, Rate$_{1}$, solver iterations, and 
Rate$_{2}$ metrics. The AI is not affected by $\alpha$ which suggests that there are
no hardware or software implementation issues, only that the Rate$_{1}$ metric tails off
as $\alpha$ is increased. We see that while the iteration growth is significant,
the Rate$_{2}$ metric is still flat for this heterogeneous and anisotropic 
steady-state diffusion problem. Thus, the primary reason that the data points 
in the static-scaling plots decrease with problem size has little to do with memory 
contention.

\section{CASE STUDY PART 1: SINGLE NODE}
\label{Sec:Studies}
The previous section, which focused entirely on the steady-state diffusion equation, 
covered the basic ways one can utilize the proposed
performance spectrum model to help justify, interpret, or diagnose the 
computational performance of any algorithm, numerical method, or solver for a
particular compute node. In these next two sections, 
we demonstrate that this performance spectrum model 
is also useful for more complicated and nonlinear PDEs. 
We consider PETSc's toy hydrostatic ice sheet 
flow example, based on the work of \citep{brown2013icesheet}, 
with geometric multigrid and apply the performance spectrum to
give us a better understanding of how certain HPC platforms scale.

\subsection{Hydrostatic ice sheet flow equations}
Consider a $[0,10]\mathrm{km} \times [0,10]
\mathrm{km} \times [0,1]\mathrm{km}$ computational ice domain $\Omega \subset \mathbb{R}^3$ lying between
a Lipschitz continuous bed $b(x,y)$ and surface $s(x,y)$. The hydrostatic equations
are obtained from the non-Newtonian Stokes equations where the horizontal 
$x-$ and $y-$ derivatives of velocity in the vertical $z$-direction are small 
and negligible. Denoting the horizontal component of the velocity field by 
$\boldsymbol{u} = (u,v)$ where $u$ and $v$ are parallel to the $x-$ and $y-$ axes
respectively, the governing equations for hydrostatic ice sheet flow is given
by
\begin{subequations}
\label{Eqn:S5_icesheet}
\begin{align}
-\eta\left(\frac{\partial}{\partial x}\left(4\frac{\partial u}{\partial x}+2\frac{\partial v}{\partial y}\right) + 
\frac{\partial}{\partial y}\left(\frac{\partial u}{\partial y}+\frac{\partial v}{\partial x}\right) +  \frac{\partial^2 u}{\partial z^2}\right) + \rho g\frac{\partial s}{\partial x}= 0\quad\mathrm{and}\\
-\eta\left(\frac{\partial}{\partial x}\left(\frac{\partial u}{\partial y}+\frac{\partial v}{\partial x}\right) +
\frac{\partial}{\partial y}\left(2\frac{\partial u}{\partial x}+4\frac{\partial v}{\partial y}\right) +  \frac{\partial^2 y}{\partial z^2}\right) + \rho g\frac{\partial s}{\partial y} = 0,
\end{align}
\end{subequations}
where $\eta$ is the nonlinear effective viscosity expressed by
\begin{align}
\eta(\gamma) = \frac{B}{2}\left(\frac{\epsilon^2}{2}+\gamma\right)^{\frac{1-n}{2n}},
\end{align}
where ice sheet models typically take $n=3$. The hardness parameter is denoted by $B$,
the regularizing strain rate is defined by $\epsilon$, and
the second invariant $\gamma$ is expressed by
\begin{align}
\gamma = \frac{\partial^2u}{\partial x^2}+\frac{\partial^2v}{\partial y^2}
+ \frac{\partial u}{\partial x}\frac{\partial v}{\partial y} + \frac{1}{4}\left(
\frac{\partial u}{\partial y}+\frac{\partial v}{\partial x}\right)^2 + \frac{1}{4}
\left(\frac{\partial^2 u}{\partial x^2}+\frac{\partial^2 v}{\partial x^2}\right).
\end{align}
More information on the theoretical derivation of the above equations can be 
found in \citep{schoof2006icestream,schoof2010thinfilm}. Equation 
\eqref{Eqn:S5_icesheet} is subject to natural boundary conditions at the free surface
and either no-slip or power-law slip conditions with friction parameter
\begin{align}
\beta^2(\gamma_b) = \beta^2_0\left(\frac{\epsilon^2_b}{2}+\gamma_b\right)^{\frac{m-1}{2}},
\end{align}
where $\gamma_b=\frac{1}{2}\left(u^2+v^2\right)$, $\epsilon_b$ is regularizing velocity,
$\beta^2_0$ is a ``low-speed" reference friction, and $m\in (0,1]$ is the slip exponent.
\subsection{Problem setup}
\begin{figure}[t]
\centering
\subfloat{\includegraphics[scale=0.42]{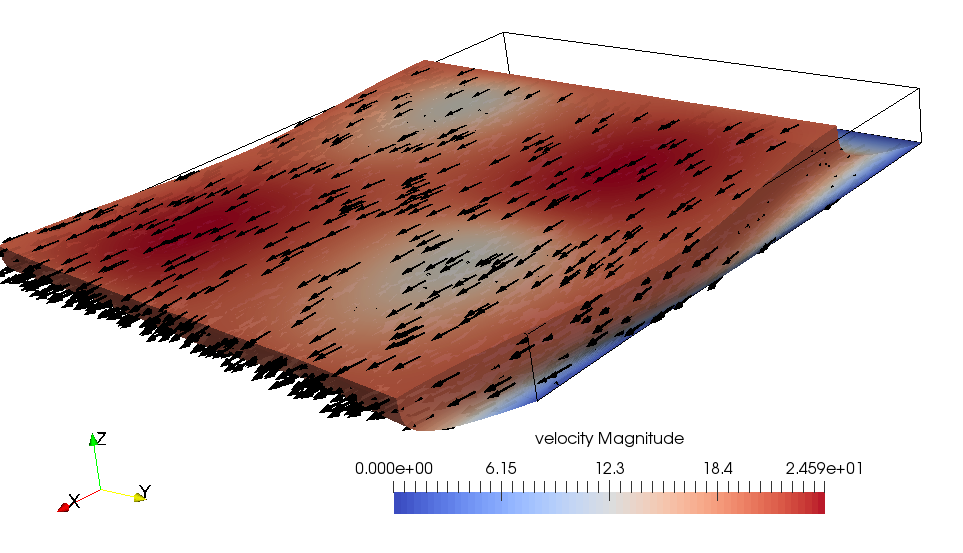}}
\captionsetup{format=hang}
\caption{Numerical solution of the velocity vector field for the 
hydrostatic ice sheet flow example.}
\label{Fig:iceflow_example}
\end{figure}

The hydrostatic ice sheet flow equation is discretized using hexahedron Q1 
finite elements on a structured grid and Figure \ref{Fig:iceflow_example}
contains the corresponding solution. Details concerning the theoretical 
derivation of the variational formulation as well as the parameters used 
for the boundary value problem can be found in \citep{brown2013icesheet}.
Since this example problem is written entirely with PETSc routines and function calls,
the [FLOPs] metric in equation \eqref{Eqn:S3_AI_simple} is determined 
using PETSc's manual FLOP counts instead of hardware FLOP counters. This is
particularly useful if a thorough comparative study on PETSc's eclectic suite
of linear algebra solvers for a particular PDE were to be conducted.

For this problem, we begin with an initial coarse grid size and successively refine
the grid $N$ times until we get the desired problem size and numerical accuracy.
The ``fine grids" produced from this element-wise refinement are solved using the geometric
multigrid technique, whereas the initial coarse grid is solved using algebraic multigrid.
The assembled Jacobian employs block AIJ format (better known as the compressed
sparse row format), where the horizontal velocity components are grouped per grid node. 
Since ice-flow is tightly coupled in the vertical direction, parallel domain decomposition 
is specially set up so that grid points in the vertical direction are never distributed
and are always contiguous in memory. The initial grid size must be chosen carefully 
because the mesh partition happens at the coarsest level and may cause load balancing 
issues if the initial grid is not large enough.
\subsection{Results}
\begin{table}
  \centering
\captionsetup{format=hang}
  \caption{Hydrostatic ice sheet flow for a single node: Mesh information and 
  number of solver iterations needed for an initial 40$\times$40$\times$5 coarse grid.
  KSP and SNES iteration counts may vary depending on the number of MPI processes used.
  \label{Tab:icesheet1}}
  \begin{tabular}{ccccc}
    \hline
    Levels of refinement & Degrees-of-freedom & SNES iterations & KSP iterations \\
    \hline
    0 & 16,000 & 7 & 39 \\
    1 & 115,200 & 8 & 45 \\
    2 & 870,400 & 8 & 44 \\
    3 & 6,758,400 & 8 & 44 \\
    \hline
  \end{tabular}
\end{table}
\begin{figure}[t]
\centering
\subfloat{\includegraphics[scale=0.5]{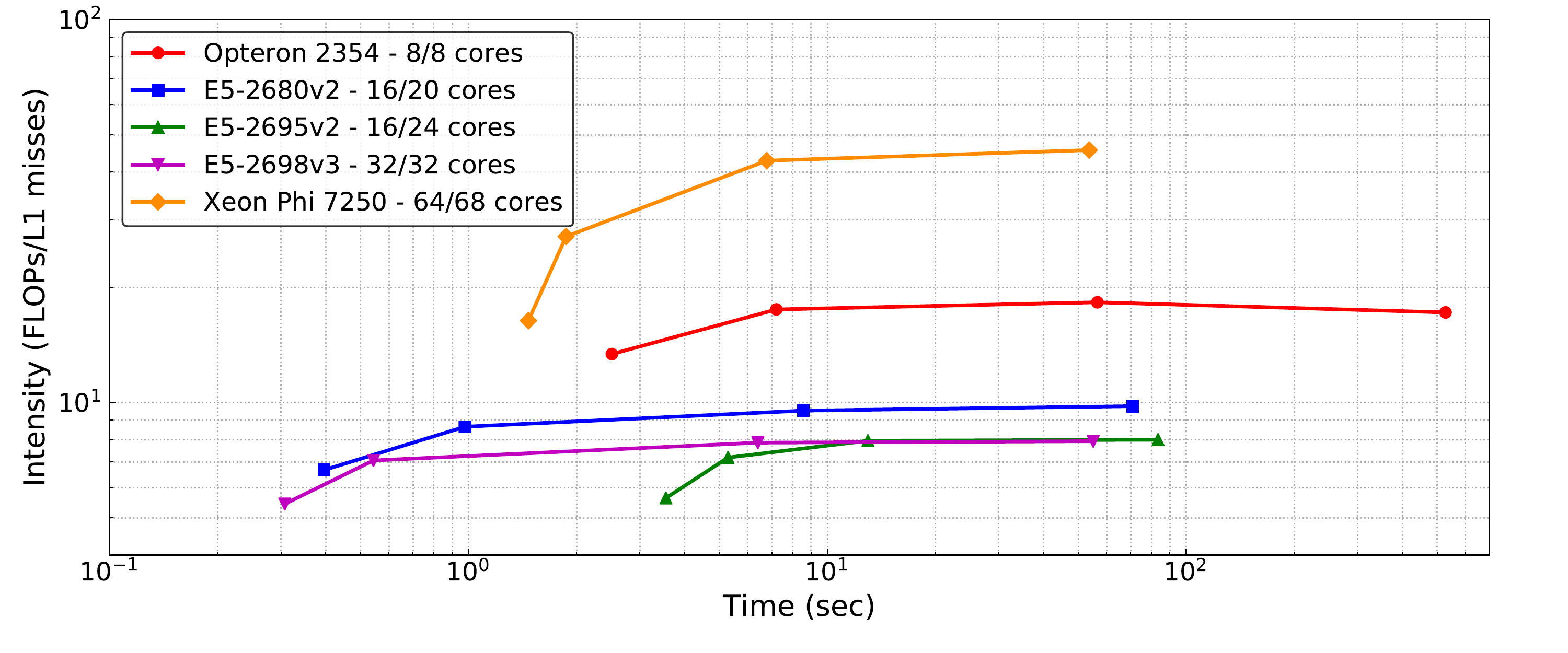}}
\captionsetup{format=hang}
\caption{Hydrostatic ice sheet flow single node: L1 arithmetic intensity, based on 
PETSc's manual FLOP counts and L1 cache misses. Note that the
two Ivybridge and KNL nodes are only partially 
saturated.}
\label{Fig:iceflow1_intensity}
\end{figure}

\begin{figure}[t!]
\centering
\subfloat{\includegraphics[scale=0.5]{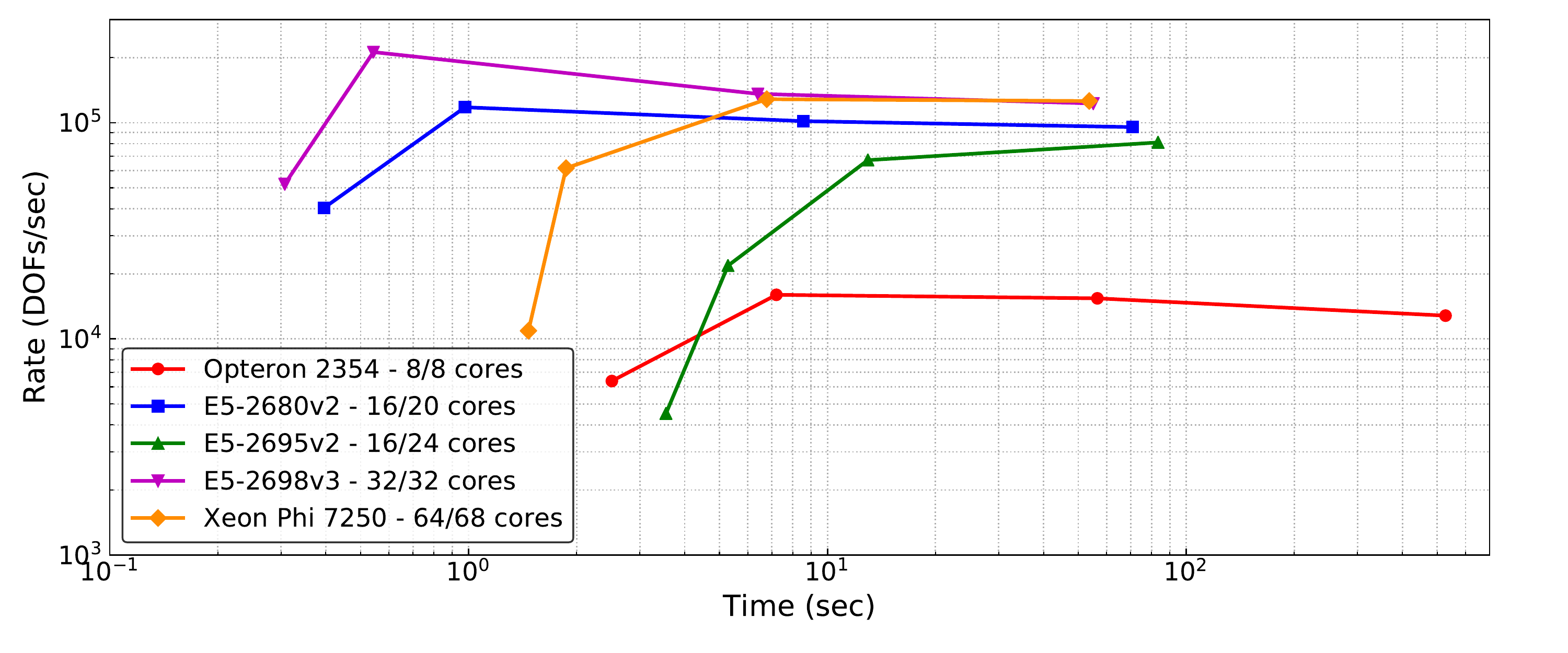}}
\captionsetup{format=hang}
\caption{Hydrostatic ice sheet flow single node: Static-scaling. 
Note that the two Ivybridge and KNL nodes are only partially saturated.}
\label{Fig:iceflow1_rate}
\end{figure}

\begin{figure}[t]
\centering
\subfloat{\includegraphics[scale=0.5]{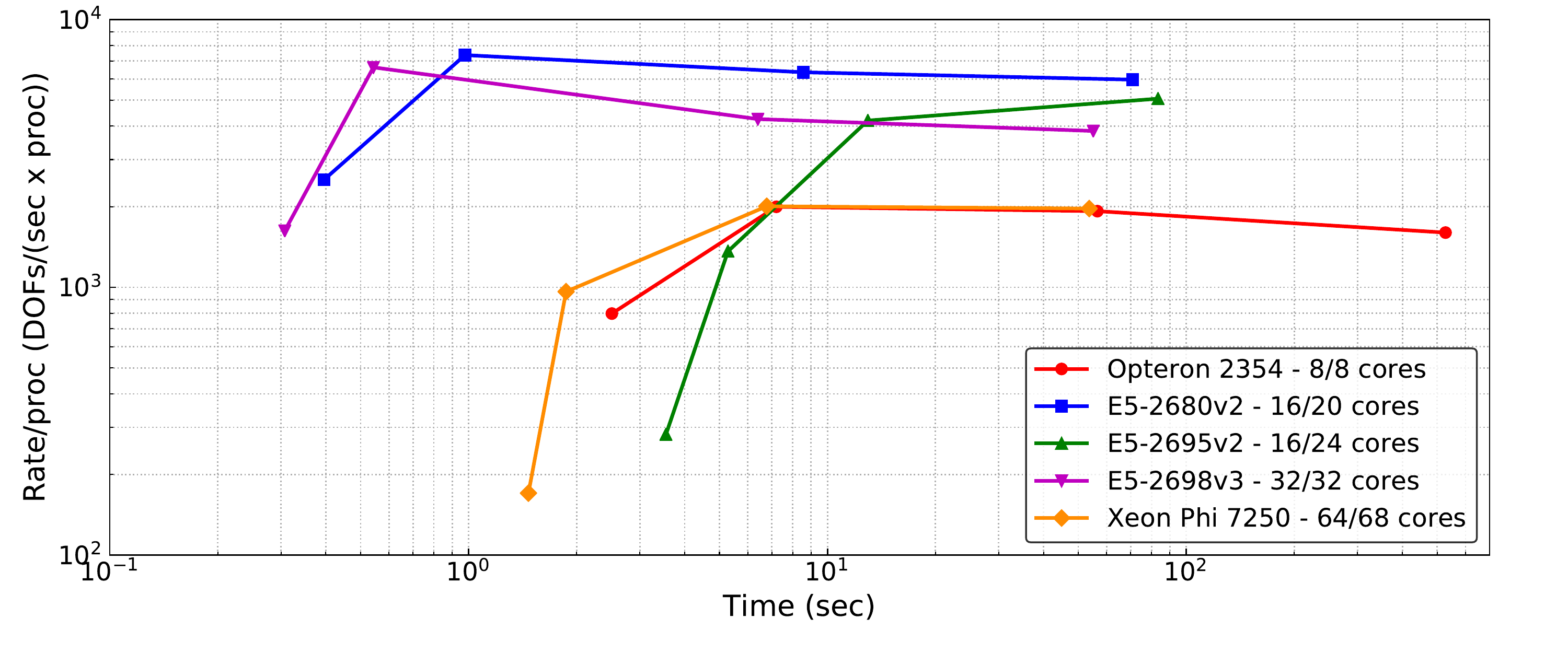}}
\captionsetup{format=hang}
\caption{Hydrostatic ice sheet flow single node: Static-scaling per MPI processes. 
Note that the two Ivybridge and KNL nodes are only partially saturated.}
\label{Fig:iceflow1_rate2}
\end{figure}
First, we provide an initial 40$\times$40$\times$5 coarse grid and 
successively refine this grid up to 3 times. All 
five processors from Table \ref{Tab:S2_HPC} are studied, and the KNL
processor is configured to use MCDRAM in flat mode. Each node has a 
different number of available cores so in order to maintain relatively 
consistent mesh partitioning, the Ivybridge and KNL processors will only
utilize 16 and 64 cores, respectively. Table \ref{Tab:icesheet1} presents the
problem size as well as the number of total SNES and KSP iterations needed 
for each level of refinement. Figure \ref{Fig:iceflow1_intensity} 
depicts the AI$_{\mbox{L1}}$ metrics with respect to the overall time-to-solution. 
Each data point has the same coarse grid size but has different levels of grid 
refinement ranging from 0 to 3. The Ivybridge'' and Haswell processors 
have similar AIs and are significantly smaller than their KNL and AMD counterparts. It should be noted
that GNU compilers were used to compile the problem on the AMD Opteron 2354 and 
Intel Xeon E5-2680v2 processors whereas the other three processors used Cray
compilers, which could explain why the AIs between the two Ivybridge processors are slightly 
different. As with the hardware counter examples in the last section, 
the AIs are initially small for the coarser
problems but eventually stabilize if a problem is sufficiently large.
It is interesting to note that the AMD processor is consistently flat for all 
data points, suggesting that the smaller problem sizes selected for this example 
have already approached the AMD processor's achievable peak performance of the
computational/memory resource. On the other hand, the KNL's wider vector 
instruction sets and caches for smaller problems are not fully utilized, 
resulting in low AI's.

The static-scaling plot on each of these compute nodes is shown in Figure 
\ref{Fig:iceflow1_rate}, and Figure \ref{Fig:iceflow1_rate2} depicts static-scaling
based on Rate$_{3}$ from \eqref{Eqn:S3_rate_mpi}. Unsurprisingly, the AMD 
processor is outperformed by all of the Intel processors. Both Haswell and KNL 
have the best performance out of all the systems studied, 
but we again notice that the KNL processor has poor metrics when the grid 
is small. Furthermore, KNL's performance per core is considerably lower as seen
from Figure \ref{Fig:iceflow1_rate2}. There are many
reasons why we noticed such dramatic behavior. First, as we already noted 
from the AI results, the problem has to be sufficiently large in order
to fully utilize the KNL vector instructions. Second, we used 64 of the 68 available
cores on the KNL node, which is at least double the amount of cores  that 
the other systems have. The degrees-of-freedom per MPI is significantly 
smaller so it is possible interprocess communication time affects the
scaling results. 

\section{CASE STUDY PART 2: MULTIPLE NODES}
\label{Sec:Studies2}
The results from every example in this paper thus far behoove us to now 
investigate what happens when more than one compute node is needed
to solve a PDE. Figures \ref{Fig:demo1_speedup} and \ref{Fig:demo2_rate}
from the previous section indicate that the AI$_{\mbox{Lx}}$ metrics
can be used to predict the strong-scaling potential on a single node. 
Our goal is now to investigate if the correlation holds true even across
multiple nodes. To ensure that the problem is sufficiently large 
to distribute to several nodes, we consider an initial 
64$\times$64$\times$6 coarse grid with three levels of refinement 
(21,495,808 degrees-of-freedom). The number of KSP and SNES iterations
needed to solve the problem are 62 and 8, respectively.
\begin{table}
  \centering
\captionsetup{format=hang}
  \caption{Hydrostatic ice sheet flow strong-scaling for an 
  initial 64$\times$64$\times$6 coarse grid.
  \label{Tab:icesheet_strong}}
  {\footnotesize
  \begin{tabular}{cccccccccc}
    \hline
    \multirow{2}{*}{Nodes} & \multicolumn{3}{|c}{E5-2695v2 (Ivybridge)} & \multicolumn{3}{|c}{E5-2698v3 (Haswell)} & \multicolumn{3}{|c}{Xeon Phi 7250 (KNL)} \\
    & \multicolumn{1}{|c}{Cores} & Time (s) & Eff. (\%) 
    & \multicolumn{1}{|c}{Cores} & Time (s) & Eff. (\%) 
    & \multicolumn{1}{|c}{Cores} & Time (s) & Eff. (\%)\\
    \hline
    1 & 16 & 300 & - & 32 & 227 & - & 64 & 193 & - \\
    2 & 32 & 150 & 100 & 64 & 108 & 105 & 128 & 92.4 & 104 \\
    4 & 64 & 72.0 & 104 & 128 & 57.3 & 99.0 & 256 & 47.3 & 102 \\
    8 & 128 & 37.9 & 98.9 & 256 & 28.7 & 98.9 & 512 & 25.2 & 95.7 \\
    16 & 256 & 18.9 & 99.2 & 512 & 13.9 & 100 & 1024 & 15.1 & 79.9\\
    32 & 512 & 9.65 & 97.2 & 1024 & 8.11 & 87.5 & 2048 & 10.6 & 56.9 \\
    64 & 1024 & 6.75 & 69.4 & 2048 &4.62 & 76.8 & 4096 & 9.27 & 32.5 \\
    \hline
  \end{tabular}}
\end{table}
\begin{table}
  \centering
\captionsetup{format=hang}
  \caption{Hydrostatic ice sheet flow for multiple nodes: Mesh information and 
  number of solver iterations needed for an initial 128$\times$128$\times$12 coarse grid.
  KSP and SNES iteration counts may vary depending
  on the number of MPI processes used.
  \label{Tab:icesheet2}}
  \begin{tabular}{ccccc}
    \hline
    Levels of refinement & Degrees-of-freedom & SNES iterations & KSP iterations \\
    \hline
    1 & 3,014,656 & 8 & 85 \\
    2 & 23,592,960 & 8 & 85 \\
    3 & 186,646,528 & 8 & 85 \\
    4 & 1,484,783,616 & 8 & 85 \\
    5 & 11,844,714,496 & 8 & 85  \\
    \hline
  \end{tabular}
\end{table}

Table \ref{Tab:icesheet_strong} contains the strong-scaling results
for the HPC systems containing the Intel Xeon E5-2695v2 (Ivybridge), Intel Xeon E5-2698v3 
(Haswell), and Intel Xeon Phi 7250 (KNL) nodes. All three systems demonstrate near
perfect strong-scaling performance until 1024 cores are used (roughly
20k degrees-of-freedom per core). However, it is difficult to make
performance comparisons because different systems employ different
numbers of MPI processes per node which affect communication to computation
ratios as well as required data bandwidth between nodes. The only concrete conclusion 
that can be made is that the KNL system takes the least amount of 
wall-clock time on a single compute node but gets outperformed when the problem
size per node reduces. Figures \ref{Fig:iceflow1_intensity} and 
\ref{Fig:iceflow1_rate} suggest that when the problem size on a KNL node
is sufficiently small, parallel performance would degrade drastically, which
is exactly what the results of Table \ref{Tab:icesheet_strong} portray.
\subsection{Example \#1: 1024 MPI processes}
\begin{figure}[t]
\centering
\subfloat{\includegraphics[scale=0.5]{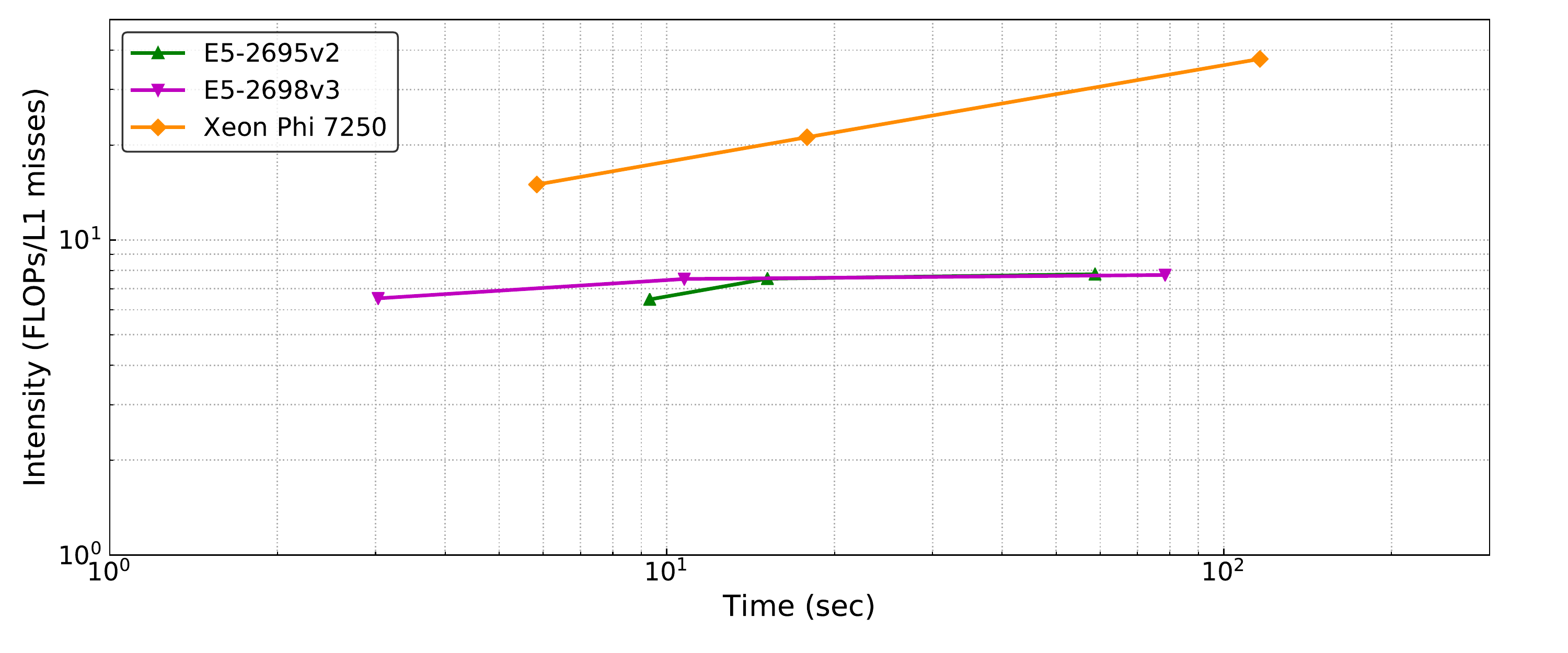}}
\captionsetup{format=hang}
\caption{Hydrostatic ice sheet flow multiple nodes: AI$_{L1}$ when the systems all employ 
1024 cores (64 Ivybridge nodes, 32 Haswell nodes, 
and 16 KNL nodes). Grid sizes ranging from 3 million to 
186 million degrees-of-freedom are considered.}
\label{Fig:iceflow2_intensity}
\end{figure}

\begin{figure}[t!]
\centering
\subfloat{\includegraphics[scale=0.5]{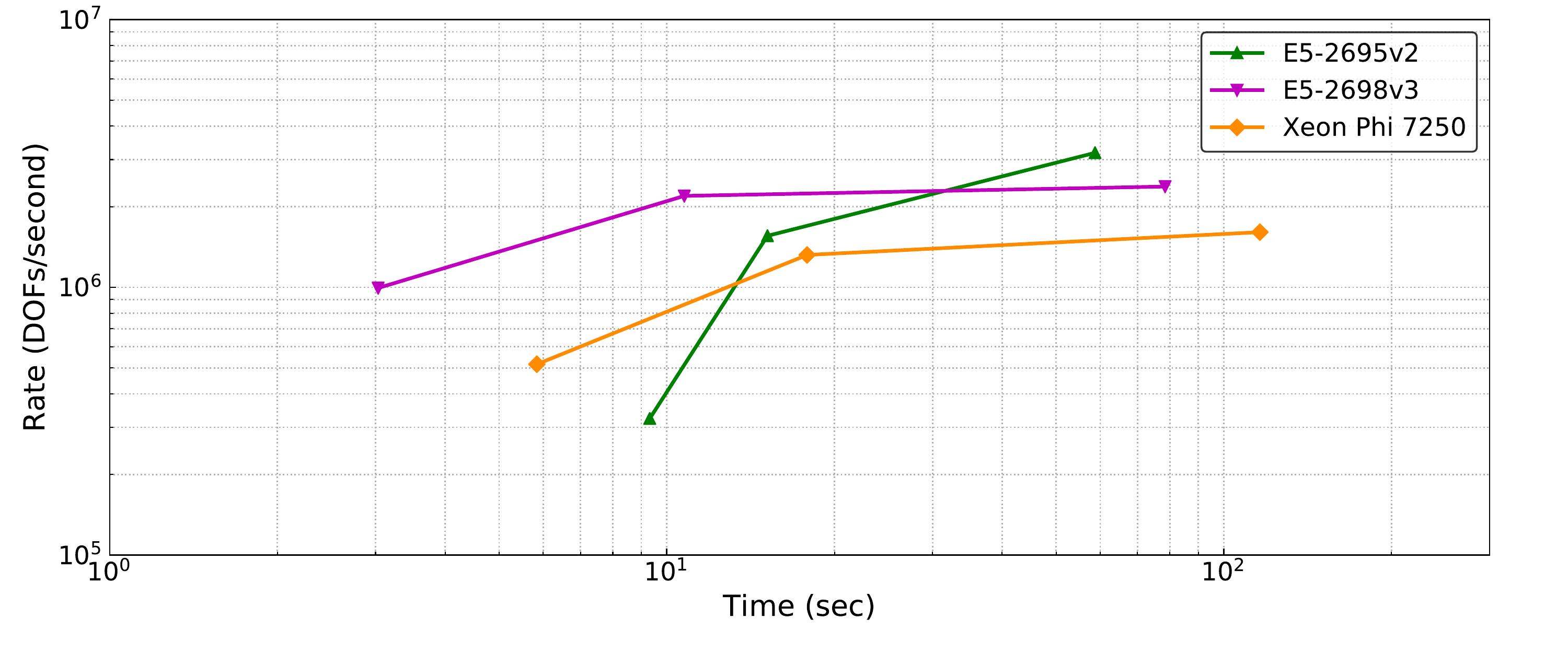}}
\captionsetup{format=hang}
\caption{Hydrostatic ice sheet flow multiple nodes: Static-scaling when 
the systems all employ 1024 MPI processes (64 Ivybridge nodes, 32 Haswell 
nodes, and 16 KNL nodes). Three levels of refinement are considered.}
\label{Fig:iceflow2_rate}
\end{figure}
\begin{figure}[t]
\centering
\subfloat{\includegraphics[scale=0.5]{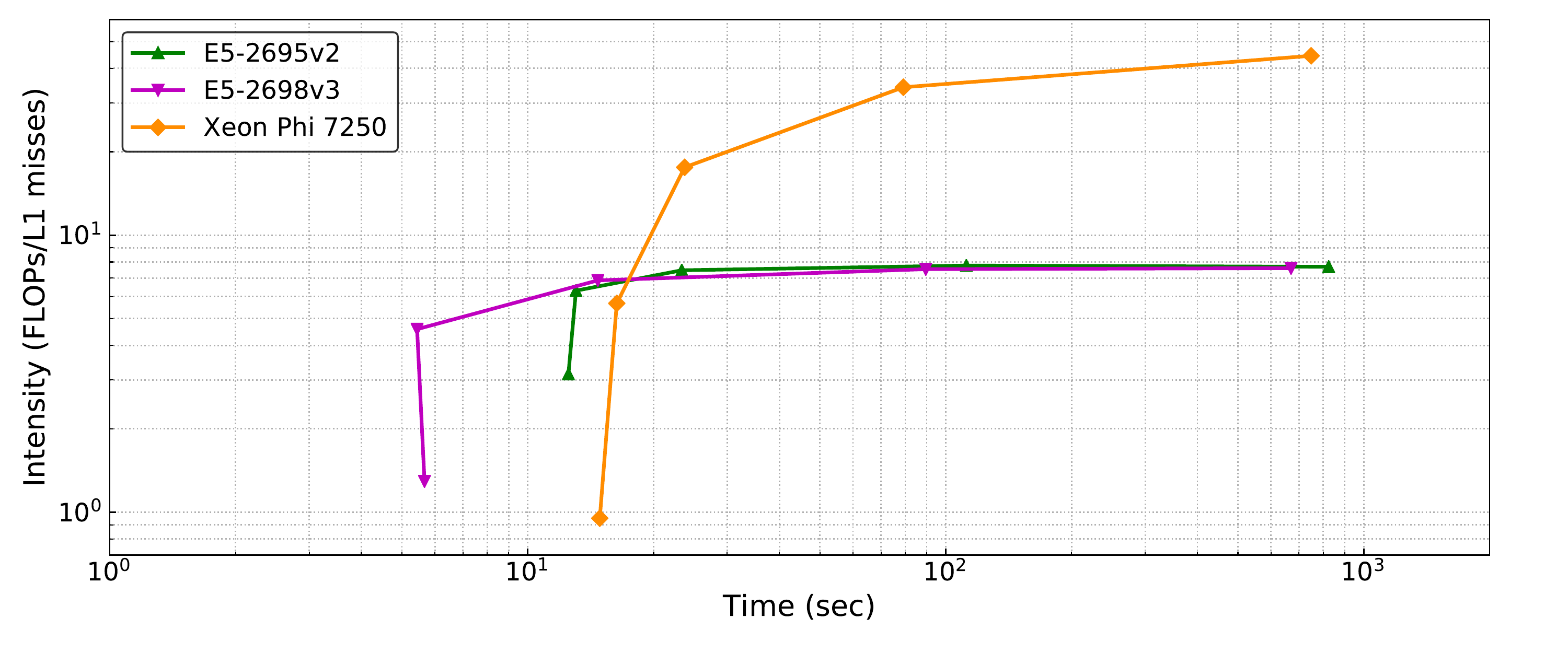}}
\captionsetup{format=hang}
\caption{Hydrostatic ice sheet flow multiple nodes: AI$_{L1}$ when the systems all employ 
256 nodes (4096, 8192, and 16384 MPI processes for Ivybridge, Haswell, and KNL 
respectively). Five levels of refinement are considered.}
\label{Fig:iceflow3_intensity}
\end{figure}
In this section, we consider what happens when we employ the same MPI concurrency. 
This second example aims to model the performance when the same hydrostatic
ice sheet flow problem is solved utilizing 1024 MPI processes on different systems. 
We set this problem up by allocating 64 Ivybridge nodes, 32 Haswell nodes, and 16 
KNL nodes. An even larger initial 128$\times$128$\times$12 
coarse grid is selected, and we refine the problem 1--3 times. 
Table \ref{Tab:icesheet2} presents the
problem size as well as the number of nonlinear and linear solver
iterations needed for every level of refinement. Figures \ref{Fig:iceflow2_intensity} 
and \ref{Fig:iceflow2_rate} contain the intensity and rate metrics, respectively. 
The AI data points are either relatively flat or do not experience drastic 
changes upon mesh refinement. The static-scaling plot tells us that the 
Ivybridge system has the best performance as the problem gets 
larger. This behavior may seem to contradict the findings of 
the static-scaling plot in Figure \ref{Fig:iceflow1_rate},
but it is important to realize that this PETSc application is limited by
the memory-bandwidth and not the TPP for the FLOP rate. The HPC
system with Ivybridge processors has the best performance simply
because it employs more compute nodes thus more available memory. 
\subsection{Example \#2: 256 compute nodes}

\begin{figure}[t]
\centering
\subfloat{\includegraphics[scale=0.5]{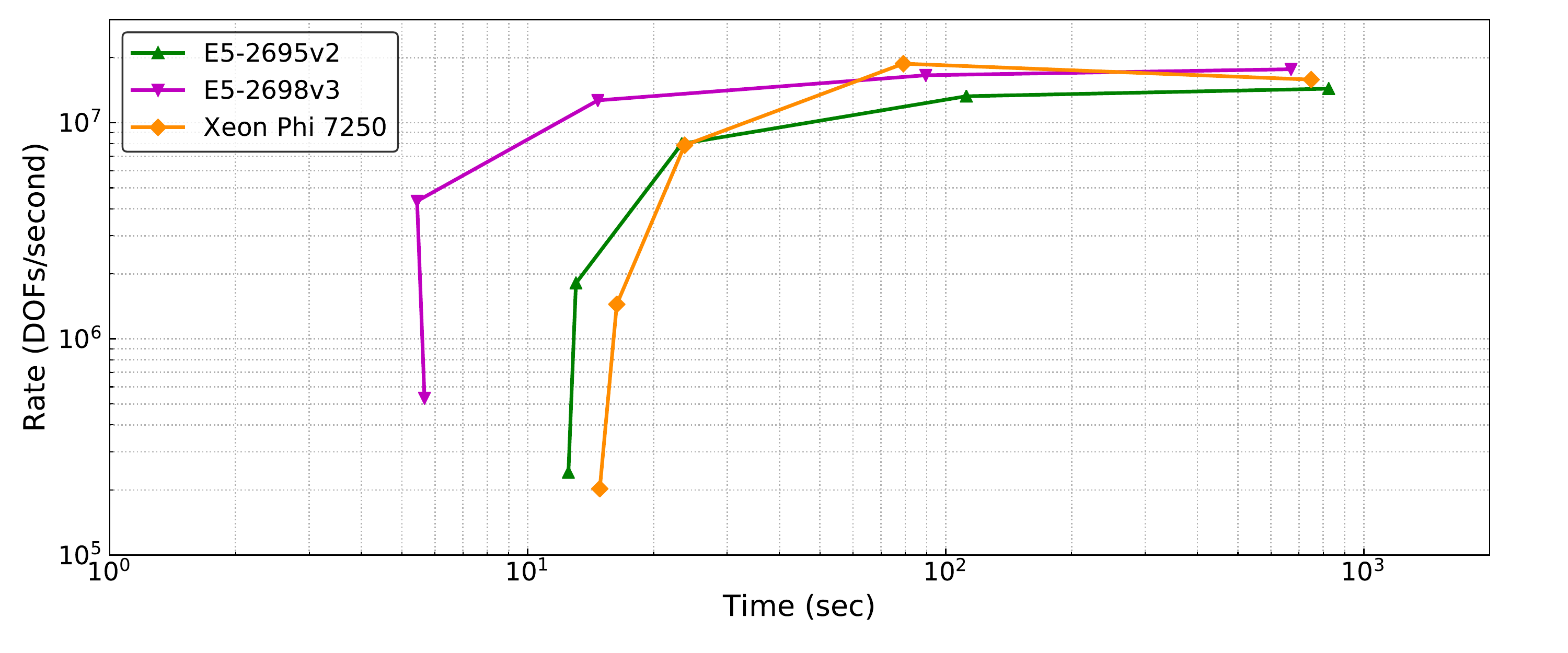}}
\captionsetup{format=hang}
\caption{Hydrostatic ice sheet flow multiple nodes: Static-scaling when 
the systems all employ 256 nodes (4096, 8192, and 16384 MPI processes 
for Ivybridge, Haswell, and KNL respectively). Five levels of refinement are
considered.}
\label{Fig:iceflow3_rate}
\end{figure}
The previous example is an elegant demonstration of why comparing
HPC machines based on equal MPI concurrency can produce misleading 
performance metrics, especially for computational frameworks that are
limited by the memory-bandwidth. What happens if every system employs 
the same number of compute nodes? In this third example, the HPC systems 
shall now allocate 256 compute nodes each. Thus, 
Ivybridge will use 4096 MPI processes (16 out of 24 cores per node), 
Haswell will use 8192 MPI processes (32 out of 32 cores per node), 
and KNL will use 16384 processes (64 out of 68 cores per node). We use the 
same initial coarse grid as in the previous example but now refine 
the problem 1--5 times. The AI in Figure 
\ref{Fig:iceflow3_intensity} again indicate relative consistency for
finer problems, and we again observe that the AI metric will 
drop significantly if a problem is not large enough. This 
trend corroborates the notion that the AI dropping for small
problems happens regardless of whether a single node or multiple nodes
are used. The static-scaling plot shown in Figure \ref{Fig:iceflow3_rate}
demonstrates that the Ivybridge processor does not beat out the 
Haswell processor. What's particularly interesting is that the performance
for the KNL processor drastically varies with problem size. KNL cannot
beat out Ivybridge for small problems, but KNL will beat both Ivybridge and Haswell
when a problem is neither too small nor too large. 

The performance spectrum model is useful for understanding performance characteristics 
across a wide variety of hardware architectures. Although the STREAM Triad
measurements from Table \ref{Tab:S2_HPC} suggest that KNL should greatly outperform
Ivybridge and Haswell for memory-bandwidth dominated applications, the performance
spectrum indicates that current and practical implementations of 
scientific software like PETSc v3.7.4 on KNL may be slow if the 
problem is dominated by main memory bandwidth. 
Different platforms require different implementation methodologies in order to
maximize performance, so optimizing computational frameworks to fully 
leverage the power of the KNL processor is still an open 
research problem. Nonetheless, the performance spectrum model
is useful for testing various implementations of PDE solvers 
and can be utilized to understand hardware architectures trends 
and algorithms of the future. 

\section{CONCLUDING REMARKS}
\label{Sec:Conclusions}
In this paper, we have proposed a performance model,
referred to as the \emph{performance spectrum}, designed
to simultaneously model both the hardware/architectural and algorithmic 
efficiencies of a variety of parallel PDE solvers. The techniques
needed to approximate such efficiency metrics are 1) the arithmetic
intensity documenting the ratio of flops over data cache misses,
and 2) static-scaling, which scales up the problem while fixing 
the concurrency. This spectrum enabled us to visualize and enrich 
our current understanding of 
performance issues related to hardware limitations, software/solver 
implementation, and numerical discretization of some popular and 
state-of-the-art finite element simulation packages and solvers. 
Moreover, it has been shown that this spectrum is also useful for 
understanding performance and scalability of complex solvers and 
PDEs for nonlinear problems like hydrostatic ice sheet flow in a 
large-scale environment. Computational scientists have designed
and are still designing software and algorithms needed to answer
many of today's pressing scientific problems, so not only do we
need to solve these problems accurately but also to solve them fast.
In order to understand how fast these solvers and software are, 
particularly ones that are either black-box or designed by others,
we need a performance model, such as the proposed performance spectrum,
to help answer any questions regarding computational performance.
\subsection{Potential extension of this work}
The performance spectrum model has been proven to be a robust and 
versatile tool for modeling the performance and scalability of 
several parallel finite element solvers and packages, but there are
still plenty of research endeavors relating to this work that need
to be addressed. We briefly provide a list of some potential
future tasks and avenues of extension:
\begin{enumerate}[leftmargin=*]
\item Most of the systems chosen for this study are from Intel and have
similar performance characteristics, but processors from other vendors such 
as IBM's POWER8 \citep{IBM_Power8}, ARM-based systems \citep{Rajovic2014322},
may tell a different story. A logical extension to this work could be to 
extend this performance spectrum analysis onto GPUs. HPC architecture is constantly 
evolving, and simple benchmarks and measurements like STREAM Triad may not 
be sufficient for understanding the performance of complex solvers or algorithms 
on these state-of-the-art machines. 
\item The intensity equations presented in Section \ref{Sec:Model} are relatively
easy to incorporate into any code, but they only provide relative comparisons 
between various flavors of hardware and software. An improved methodology for 
documenting the [Work] and [TBT] metrics would certain improve the predictive
capabilities of the performance spectrum model. Other popular techniques such
as using Intel's SDE and VTune libraries can be used to measure AI on CPUs.
\item Certain PDEs, like the advection-diffusion equation, are notoriously difficult
to solve especially for high P\'eclet numbers. The performance spectrum can be useful 
to compare not only various numerical discretization (e.g., finite element vs
finite volume methods) but also special solvers when physical properties such as
velocity and diffusivity vary with length scale. Moreover, complex PDEs may also involve
multiple physical components, thus requiring mixed formulations or hybridization
techniques that may not be so trivial to solve or implement.
\item Finally, the performance spectrum is not restricted to solving PDEs. One 
can examine any computational scientific problem where performance and scalability
is limited by problem size. Examples include but are not limited to: boundary integral
equations, uncertainty quantification, parameter estimation, adaptive mesh refinement,
and parallel domain decomposition. 
\end{enumerate}

\section*{ACKNOWLEDGMENTS}
The authors acknowledge Jed Brown (University of Colorado at Boulder) 
for his invaluable thoughts and ideas which inspired the creation of this paper. 
Compute time on the Maxwell and Opuntia systems is provided
by the Center for Advanced Computing \& Data Systems (CACDS) at the 
University of Houston, and compute time on the Edison and Cori systems is
provided by the National Energy Research Scientific Computing Center (NERSC).
MGK acknowledges partial support from the Rice Intel Parallel Computing Center and US DOE Contract DE-AC02-06CH11357.
The opinions expressed in this paper are those of the authors and 
do not necessarily reflect that of the sponsors. 

\bibliographystyle{plainnat}
\bibliography{references}
\end{document}